\pdfoutput=1
\documentclass[a4paper,11pt]{article}
\usepackage{jheppub}
\usepackage{amsmath,amssymb}
\usepackage{hyperref}
\usepackage{physics}
\usepackage{float}

\newcommand{\hmntt}[0]{h_{\mu \nu}^{TT}}
\newcommand{\TTproj}[0]{\varepsilon_\mn^{TT,\ab}}

\newcommand{\sfX}[0]{\mathsf{X}}
\newcommand{\sfY}[0]{\mathsf{Y}}
\newcommand{\sfM}[0]{\mathsf{M}}
\newcommand{\sfZ}[0]{\mathsf{Z}}

\newcommand{\ab}[0]{{\alpha \beta}}
\newcommand{\mn}[0]{{\mu \nu}}
\newcommand{\MN}[0]{{M N}}
\renewcommand{\ij}[0]{{ij}}
\newcommand{\pd}[0]{\partial}

\newcommand{\B}[0]{\mathcal{B}}
\newcommand{\C}[0]{\mathcal{C}}
\newcommand{\D}[0]{\mathcal{D}}
\newcommand{\E}[0]{\mathcal{E}}
\newcommand{\F}[0]{\mathcal{F}}
\newcommand{\J}[0]{\mathcal{J}}
\renewcommand{\L}[0]{\mathcal{L}}
\newcommand{\N}[0]{\mathcal{N}}
\renewcommand{\O}[0]{\mathcal{O}}
\newcommand{\R}[0]{\mathcal{R}}
\renewcommand{\S}[0]{\mathcal{S}}
\newcommand{\T}[0]{\mathcal{T}}
\newcommand{\W}[0]{\mathcal{W}}

\newcommand{\oford}[1]{\mathcal{O}\qty(#1)}
\newcommand{\liegroup}[3]{\text{#1}\qty(#2)^{#3}}

\renewcommand{\dd}{\mathrm{d}}

\newcommand{\alphasymbol}[0]{\Omega}

\makeatletter
\def\@fpheader{\relax}
\makeatother

\preprint{\text{HIP-2026-23/TH}}

\title{Gravitational form factors of \texorpdfstring{$\rho$ and $a_1$}{rho and a1} mesons in V-QCD}

\author[a]{Antti Hippel\"ainen,}
\affiliation[a]{Department of Physics and Helsinki Institute of Physics, \\
P.O. Box 64, FI-00014, University of Helsinki, Finland}

\emailAdd{antti.hippelainen@helsinki.fi}

\abstract{We compute the gravitational form factors and mechanical properties of the $\rho$ and $a_1$ mesons in V-QCD, a holographic model of QCD with backreacting flavor and dynamical chiral symmetry breaking. The parameters of the model are taken from an earlier fit to meson spectra and are not adjusted to form factor data. Both mesons are described by a single spin-1 field, and the six form factors are identified from the matrix element of the energy-momentum tensor with two spin-1 states, following the holographic dictionary. We find that $A_0 = J$ and $A_1 = 0$ hold identically in the model, and show that the remaining form factors $E$, $D_0$ and $D_1$ are regular at zero momentum transfer. We compare the form factors of $\rho$ with results from the Nambu--Jona-Lasinio model, a light-cone constituent quark model and the hard-wall holographic model. From the form factors, we compute the monopole and quadrupole distributions of energy, pressure and shear force in the Breit frame. The D-term, defined for spin-1 hadrons as the forward value of the combination $\D_0(t)$ of form factors, is found to be $-2.98$ for the $\rho$ and $-7.43$ for the $a_1$ meson. The mass and mechanical radii are $0.81$~fm and $0.89$~fm for $\rho$, and $0.73$~fm and $0.80$~fm for $a_1$.}

\begin{document}
\maketitle
\flushbottom
\setcounter{page}{2}

\clearpage

\section{Introduction}
\label{sec:intro}

The gravitational form factors (GFFs) of a hadron describe how mass, spin and internal forces are distributed inside the particle~\cite{Polyakov:2002yz, Polyakov:2018zvc}. The GFFs parametrize the matrix element of the energy-momentum tensor (EMT) between hadron states, in the same way as the electromagnetic form factors parametrize the matrix elements of the electromagnetic current. Most studies so far have focused on the nucleon and the pion, whose EMT matrix elements contain three and two form factors, respectively. The number of independent structures grows with the spin of the hadron~\cite{Cotogno:2019vjb}, and for a spin-1 hadron the matrix element of the total EMT contains six form factors~\cite{Holstein:2006ud, Cosyn:2019aio, Polyakov:2019lbq}. Two of the values of these GFFs at zero momentum transfer are fixed by Poincar\'e symmetry~\cite{Lorce:2019sbq, Cotogno:2019xcl}, while the other four are dynamical. 

The lightest spin-1 hadrons are the $\rho$ meson and its chiral partner, the $a_1$ meson. For $\rho$, the GFFs have been computed in the Nambu--Jona-Lasinio (NJL) model~\cite{Freese:2019bhb}, in a light-cone constituent quark model~\cite{Sun:2020wfo}, from light-front wave functions~\cite{Shi:2023oll}, with light-cone QCD sum rules~\cite{Aliev:2020aih}, in chiral effective field theory~\cite{Epelbaum:2021ahi}, and holographically in the hard-wall~\cite{Abidin:2008ku} and soft-wall~\cite{Allahverdiyeva:2023fhn} models. The D-term of $\rho$ has also been studied in the bag model~\cite{Neubelt:2019sou} and with dilaton effective theory~\cite{Stegeman:2025tdl}. For $a_1$, the only computation we are aware of is the determination of the tensor contribution to the form factors in the hard-wall model~\cite{Abidin:2008hn}. In lattice QCD, the quark momentum fraction of $\rho$ was computed in the quenched approximation~\cite{Best:1997qp}, and the only determination of the $t$-dependence of the form factors of $\rho$ that we are aware of is that of the gluon contributions at a pion mass $m_\pi = 450$~MeV~\cite{Pefkou:2021fni}. The different model determinations differ considerably, already in the forward values of the form factors.

In this work we compute the GFFs of the $\rho$ and $a_1$ mesons in V-QCD~\cite{Jarvinen:2011qe}, a bottom-up holographic model of QCD in the Veneziano limit, in which the glue and flavor sectors are both dynamical and backreact on each other, and chiral symmetry is broken dynamically by a tachyonic scalar in the flavor sector. We use the potentials of~\cite{Amorim:2021gat}, fitted to meson spectra, and the same background as in the computation of the GFFs of the pion in~\cite{Hippelainen:2026pion}, which extended the computation for the proton in improved holographic QCD~\cite{Hippelainen:2026izm}. The $\rho$ and $a_1$ are identified as the lowest normalizable modes of non-singlet vector and transverse axial-vector fields~\cite{Arean:2013tja}, whose actions differ only by a mass term induced by the tachyon. We can therefore treat both mesons in a unified manner. The matrix element of the EMT is obtained with LSZ reduction, so that by the holographic dictionary, the matrix element is computed from the cubic couplings of the unified field to the tensor and scalar fluctuations of the metric. The scalar fluctuations of the metric mix with those of the dilaton and the tachyon. The tensor channel determines four of the six form factors, and the scalar channel with the tensor channel the two remaining ``scalar'' form factors, $D_0$ and $D_1$.

We find that in this model, $A_0 = J$ and $A_1 = 0$ hold identically for both mesons, since the cubic vertices following from the flavor action do not generate the tensor structure corresponding to the one multiplying $A_1$ in the decomposition. The forward value of $E$ is below its free-field value of one~\cite{Polyakov:2019lbq}, $E(0) = 0.517$ for $\rho$ and $0.430$ for $a_1$. The expressions for $E$, $D_0$ and $D_1$ contain explicit inverse powers of $t$, and we show that all three form factors are nevertheless regular at $t = 0$. From the form factors we compute the Breit-frame distributions of energy, pressure and shear force. A polarized spin-1 hadron is not spherically symmetric, so in addition to the monopole distributions of energy, pressure and shear force, a spin-1 particle also has quadrupole distributions~\cite{Polyakov:2019lbq}. The D-term, defined for spin-1 hadrons as the forward value of the combination $\D_0$ of the form factors, is found to be negative, $-2.98$ for $\rho$ and $-7.43$ for $a_1$. For both mesons, we find that the mass and mechanical radii are of the same order, $0.81$~fm and $0.89$~fm for $\rho$, and $0.73$~fm and $0.80$~fm for $a_1$, respectively. 

The rest of the paper is organized as follows. In Section~\ref{sec:holographicmodel} we present the holographic model, the gravity fluctuations and the identification of vector and axial-vector mesons. In Section~\ref{sec:gravitationalformfactors} we derive the matrix element of the energy-momentum tensor between spin-1 states and identify the six form factors from the tensor and scalar vertices. In Section~\ref{sec:results} we present the form factors, compare them with other determinations, and compute the Breit-frame distributions, the D-terms and the radii. We conclude in Section~\ref{sec:conclusions}. The boundary values of the gravity fluctuations are discussed in Appendix~\ref{app:boundaryvalues}, the cubic vertices are derived in Appendix~\ref{subapp:quadraticactionandcubicvertices}, and the regularity of the form factors at zero momentum transfer is shown in Appendix~\ref{subapp:regularity}.

\section{Holographic model}
\label{sec:holographicmodel}
We analyze the structure of the $\rho$ and $a_1$ mesons in V-QCD~\cite{Jarvinen:2011qe}, a five-dimensional holographic model of QCD. The background and the scalar sector of the gravity fluctuations are the same as in~\cite{Hippelainen:2026pion}, and we summarize them to the extent needed below. Two ingredients are new: the transverse vector modes of the metric fluctuation, which we keep explicit, and the vector and axial-vector mesons themselves. Section~\ref{subsec:vqcdandthebackground} defines the model and its background, Section~\ref{subsec:gravityfluctuations} the gravity fluctuations, and Section~\ref{subsec:vectorandaxialvectormesons} the $\rho$ and $a_1$ modes.

\subsection{V-QCD and the background}
\label{subsec:vqcdandthebackground}

V-QCD is a bottom-up holographic model of QCD in the Veneziano limit, where the number of colors $N_c$ and flavors $N_f$ are taken to infinity, and the ratio $x \equiv N_f/N_c$ is kept constant~\cite{Jarvinen:2011qe}. We set $x = \frac{2}{3}$, corresponding to two active quark flavors and $N_c = 3$, the values used in the fit of the model potentials in~\cite{Amorim:2021gat}. The glue sector of V-QCD is a version of improved holographic QCD~\cite{Gursoy:2007cb, Gursoy:2007er}. With it, dynamical flavor is included by $N_f$ pairs of space-filling $\text{D4}$--$\overline{\text{D4}}$ branes with an open-string tachyon, whose dynamics is described by a tachyonic Dirac--Born--Infeld (DBI) action. In the Veneziano limit the flavor sector backreacts at leading order.

The fields of the model are the metric, dual to the energy-momentum tensor $T^\mn$, the dilaton $\Phi$, dual to $\Tr G_\mn G^\mn$, where $G_\mn$ is the field-strength tensor of gluons, the tachyon $T^{ij}$, dual to the quark bilinear $\bar{q}^i_R q^j_L$, and the $\liegroup{U}{N_f}{}_L \times \liegroup{U}{N_f}{}_R$ gauge fields $L_M^{ij}$ and $R_M^{ij}$, dual to the left- and right-handed flavor currents~\cite{Jarvinen:2011qe}. The exponential $e^\Phi = \lambda$ of the dilaton is identified with the 't Hooft coupling~\cite{Gursoy:2007cb}. Capital Latin indices $M, N, \dots$ run over the five bulk coordinates, Greek indices $\mu, \nu, \dots$ over the four boundary coordinates, $i, j = 1, \dots, N_f$ are here fundamental flavor indices, and $a, b, \dots = 1,\dots, N_f^2-1$ are adjoint flavor indices. The flavor group generators $t^a$ are normalized as $\Tr(t^a t^b) = \frac{1}{2}\delta^{ab}$, and a field in the adjoint representation is expanded as $X^{ij} = X^a (t^a)^{ij}$. From here on we suppress the fundamental indices and write such fields as $X = X^a t^a$. Repeated indices are summed over unless stated otherwise, signature is mostly plus with $\eta_\mn = \text{diag}(-1, 1, 1, 1)$, and $\Box = \eta^\mn \pd_\mu \pd_\nu$.

The action is~\cite{Jarvinen:2011qe}
\begin{equation}
    \label{eq:vqcdaction}
    \begin{split}
        S &= S_g + S_f \ , \\
        S_g &= M^3 N_c^2 \int \dd^5 x \sqrt{-g}\qty[\R - \frac{4}{3} (\pd \Phi)^2 + V_g(\Phi)] \ , \\
        S_f &= -\frac{1}{2} M^3 N_c \Tr \int \dd^5 x \qty[V_f\qty(\Phi, T^\dagger T)\sqrt{-\det \boldsymbol{L}} + V_f\qty(\Phi, T T^\dagger)\sqrt{-\det \boldsymbol{R}}] \ ,
    \end{split}
\end{equation}
where $M$ is the five-dimensional Planck mass, $\sqrt{-g}$ the square root of the determinant of the metric, $\R$ the Ricci scalar, and $\Tr$ denotes the symmetrized trace over (suppressed) flavor indices, reducing to the ordinary trace at quadratic order in flavor fluctuations, at which we work. The DBI matrices are
\begin{equation}
    \label{eq:dbimatrices}
    \begin{split}
        \boldsymbol{L}_\MN &= g_\MN + w(\Phi) F^L_\MN + \frac{\kappa(\Phi)}{2}\qty[(D_M T)^\dagger (D_N T) + (D_N T)^\dagger (D_M T)] \ , \\
        \boldsymbol{R}_\MN &= g_\MN + w(\Phi) F^R_\MN + \frac{\kappa(\Phi)}{2}\qty[(D_M T)(D_N T)^\dagger + (D_N T)(D_M T)^\dagger] \ ,
    \end{split}
\end{equation}
with the covariant derivative and field strengths
\begin{equation}
    \label{eq:covariantderivative}
    D_M T = \pd_M T + i T L_M - i R_M T \ , \qquad F^L_\MN = \pd_M L_N - \pd_N L_M - i\comm{L_M}{L_N} \ ,
\end{equation}
and $F^R_\MN$ analogously. The potentials $V_g$, $V_f$, $\kappa$ and $w$ are those of~\cite{Amorim:2021gat}, where their parameters were fitted to meson spectra. They are given explicitly, in the conventions used here, in~\cite{Hippelainen:2026pion}. Of the potentials, $V_g$ depends on $\Phi$ alone and $V_f$ on both $\Phi$ and the tachyon, while for $\kappa$ and $w$ we have left out a possible dependence on the tachyon. V-QCD also contains a CP-odd sector~\cite{Arean:2013tja} with an axion, but it does not contribute to the analysis in this work, and is hence omitted.

For the background we use the Ansatz
\begin{equation}
    \label{eq:backgroundansatz}
    \dd s^2 = e^{2A(z)}\qty(\eta_\mn \dd x^\mu \dd x^\nu + \dd z^2) \ , \qquad \Phi = \Phi(z) \ , \qquad T = \tau(z) \mathbf{1}_{N_f} \ ,
\end{equation}
with vanishing gauge fields, the UV boundary at $z = 0$ and an IR singularity at $z \to \infty$. A prime denotes $\pd_z$ throughout, and unless required for clarity, explicit dependence of fields on the holographic bulk coordinate $z$ is suppressed from now on. On the background, both $\boldsymbol{L}$ and $\boldsymbol{R}$ reduce to an effective metric
\begin{equation}
    \label{eq:effectivemetric}
    \widetilde{g}_\MN = g_\MN + \kappa(\Phi) \pd_M \tau \pd_N \tau = e^{2A} \text{diag}\qty(\eta_\mn, G^2) \ , \qquad G = \sqrt{1 + e^{-2A}\kappa (\tau')^2} \ .
\end{equation}
The background is solved from the Einstein equations and the equations of motion of $\Phi$ and $\tau$, which read~\cite{Jarvinen:2011qe}
\begin{equation}
    \label{eq:backgroundeoms}
    \begin{split}
        3\qty(A'' + (A')^2) &= - \frac{2}{3} \qty(\Phi')^2 + \frac{1}{2} e^{2A}V_g - \frac{1}{2} x V_f e^{2A} G \ , \\
        6 \qty(A')^2 &= \frac{2}{3}\qty(\Phi')^2 + \frac{1}{2}e^{2A}V_g - \frac{1}{2}x V_f \frac{e^{2A}}{G} \ , \\
        0 &= \Phi'' + 3 A' \Phi' + \frac{3}{8}e^{2A} \qty(\pd_\Phi V_g - x G \pd_\Phi V_f - e^{-2A} x V_f (\pd_\Phi \kappa) \frac{\qty(\tau')^2}{2G}) \ , \\
        0 &= e^{5A} G \pd_\tau V_f - \pd_z \qty(e^{3A} V_f \kappa \frac{\tau'}{G}) \ .
    \end{split}
\end{equation}
The difference of the first two equations is used repeatedly below,
\begin{equation}
    \label{eq:eeconstraint}
    A'' - \qty(A')^2 = - \frac{4}{9}\qty(\Phi')^2 - \frac{x}{6}\frac{V_f \kappa \qty(\tau')^2}{G} \ .
\end{equation}
We solve~\eqref{eq:backgroundeoms} numerically with the method of~\cite{Amorim:2021gat}. Regularity in the IR leaves one free parameter in the tachyon solution, which sets the quark mass, and we choose it such that the lowest pseudoscalar state has the physical pion mass, $m_\pi \approx 135$~MeV~\cite{Hippelainen:2026pion}. All dimensionful quantities of the model are given in units of the only dimensionful parameter of the model, $\Lambda_\text{UV}$, which is fixed by setting the mass of the lowest vector state to the mass of the $\rho$ meson, $m_\rho = 775.3$~MeV, as in~\cite{Amorim:2021gat}. All model parameters are thereby fixed by meson spectra, and none are adjusted to form factor data.

\subsection{Gravity fluctuations}
\label{subsec:gravityfluctuations}

Here we parametrize the metric, dilaton and tachyon fluctuations, organize them into gauge-invariant combinations, and define the bulk-to-boundary propagators needed for the vertices in Section~\ref{sec:gravitationalformfactors}. The energy-momentum tensor is sourced by the boundary value of the metric, so we must fluctuate with respect to the metric as well as the dilaton and the tachyon. We parametrize the fluctuated metric as in~\cite{Kiritsis:2006ua, Arean:2013tja}, keeping also the transverse vector parts,
\begin{equation}
    \label{eq:fluctuatedmetric}
    \begin{split}
        \dd s^2 = e^{2A}\Big[&\qty(1 + 2\phi)\dd z^2 + 2\qty(\pd_\mu W + A^T_\mu)\dd z \dd x^\mu \\
        &+ \qty(\qty(1 + 2\psi)\eta_\mn + 2\pd_\mu \pd_\nu E + \pd_{\{\mu} V^T_{\nu\}} + 2 h^{TT}_\mn)\dd x^\mu \dd x^\nu\Big] \ ,
    \end{split}
\end{equation}
where the braces denote symmetrization, $\pd_{\{\mu} V^T_{\nu\}} = \pd_\mu V^T_\nu + \pd_\nu V^T_\mu$, together with $\Phi(x^M) = \Phi(z) + \chi(x^M)$ and $T(x^M) = \qty(\tau(z) + s(x^M))\mathbf{1}_{N_f}$. All fluctuations depend on $x^M$ and are infinitesimal, the vectors are transverse, $\pd^\mu A^T_\mu = \pd^\mu V^T_\mu = 0$, and $h^{TT}_\mn$ is transverse and traceless. The boundary metric is a flavor-singlet and parity-even source, so at linear order it does not couple with the flavor non-singlet or the pseudoscalar fluctuations of the tachyon, and we omit them.

Not all of the fluctuation fields introduced above are physical, since some of them can be removed by diffeomorphisms. Under an infinitesimal bulk diffeomorphism $x^\mu \mapsto x^\mu + \xi^\mu$, $z \mapsto z + \xi_z$, with $\xi_\mu = \pd_\mu \xi + \xi^T_\mu$ and $\pd^\mu \xi^T_\mu = 0$, the fluctuations transform as
\begin{equation}
    \label{eq:diffeotransformations}
    \begin{split}
        &\psi \mapsto \psi - A' \xi_z \ , \qquad \chi \mapsto \chi - \Phi' \xi_z \ , \qquad s \mapsto s - \tau' \xi_z \ , \\
        &\phi \mapsto \phi - \xi_z' - A' \xi_z \ , \qquad W \mapsto W - \xi_z - \xi' \ , \qquad E \mapsto E - \xi \ , \\
        &A^T_\mu \mapsto A^T_\mu - \qty(\xi^T_\mu)' \ , \qquad V^T_\mu \mapsto V^T_\mu - \xi^T_\mu \ .
    \end{split}
\end{equation}
Since there are six scalars and two scalar gauge functions $\xi_z$ and $\xi$, we can form four functionally independent diffeomorphism-invariant scalars. Similarly, the two transverse vectors and the transverse gauge function $\xi^T_\mu$ leave one invariant vector. We write the vertices in terms of the following invariants and choose $\qty{\chi, E, V^T_\mu}$ to be pure gauge,
\begin{equation}
    \label{eq:gaugeinvariants}
    \begin{split}
        &\zeta = \psi - \frac{A'}{\Phi'}\chi \ , \quad \varsigma = \psi - \frac{A'}{\tau'}s \ , \quad \F = \frac{\psi}{A'} - \qty(W - E') \ , \\
        &\Upsilon = \phi - \pd_z \qty(\frac{\psi}{A'}) - \psi \ , \quad \W^T_\mu = A^T_\mu - (V_\mu^T)' \ .
    \end{split}
\end{equation}
Here $\zeta$ and $\varsigma$ are the invariants of~\cite{Arean:2013tja}, while $\F$ and $\Upsilon$ generalize the corresponding combinations of improved holographic QCD~\cite{Kiritsis:2006ua, Hippelainen:2026izm}.

The $(\mu z)$ components of the linearized Einstein equations, $\psi' - A'\phi + \frac{4}{9}\Phi'\chi + \frac{x V_f \kappa \tau'}{6G}s = 0$, imply that on-shell $\Upsilon$ is not independent of the other invariants~\cite{Arean:2013tja},
\begin{equation}
    \label{eq:upsilonrelation}
    \Upsilon = - \frac{4\qty(\Phi')^2}{9\qty(A')^2}\zeta - \frac{x V_f \kappa \qty(\tau')^2}{6G\qty(A')^2}\varsigma \ ,
\end{equation}
with $G$ defined in~\eqref{eq:effectivemetric}, so $\Upsilon$ can be eliminated.

The three remaining scalar invariants are also related on-shell. The part of the $(\mu\nu)$ components of the linearized Einstein equations that is proportional to $\pd_\mu \pd_\nu$ reads $\phi + 2\psi = (W - E')' + 3A'(W - E')$~\cite{Arean:2013tja}. Differentiating the definition of $\F$ in~\eqref{eq:gaugeinvariants} and using this equation gives $\F' + 3A'\F = -\Upsilon$, which with~\eqref{eq:upsilonrelation} becomes
\begin{equation}
    \label{eq:conjugateequation}
        \F' + 3A' \F = \frac{4\qty(\Phi')^2}{9\qty(A')^2}\zeta + \frac{x V_f \kappa \qty(\tau')^2}{6G\qty(A')^2}\varsigma \ .
\end{equation}
This relation determines $\F$ once $\zeta$ and $\varsigma$ are known.

For the vertex computations of Section~\ref{sec:gravitationalformfactors} we need the original fluctuations in terms of the invariants. Inverting~\eqref{eq:gaugeinvariants} gives
\begin{equation}
    \label{eq:invertedscalars}
    \begin{split}
        \psi &= \zeta + A' \frac{\chi}{\Phi'} \ , \quad s = \frac{\tau'}{A'}(\zeta - \varsigma) + \tau' \frac{\chi}{\Phi'} \ , \quad W = E' + \qty(\frac{\psi}{A'} - \F) \ , \\
        \phi &= \frac{\psi'}{A'} + \frac{4\qty(\Phi')^2}{9(A')^2}(\psi - \zeta) + \frac{x V_f \kappa \qty(\tau')^2}{6G(A')^2}(\psi - \varsigma) \ , \quad A_\mu^T = \W_\mu^T + \qty(V_\mu^T)' \ ,
    \end{split}
\end{equation}
where the gauge degrees of freedom $\qty{\chi, E, V_\mu^T}$ remain. They must drop out of the vertices, which we show explicitly in Appendix~\ref{subapp:gaugeinvariance}.

We denote the Fourier transform of a quantity defined on the boundary by a tilde, and that of a field depending also on $z$ by a hat,
\begin{equation}
    \label{eq:fourierconvention}
    \widetilde{f}(q) = \int \dd^4 x e^{-iq\cdot x} f(x) \ , \qquad \widehat{f}(q, z) = \int \dd^4 x e^{-iq\cdot x} f(x, z) \ .
\end{equation}
Since the background depends only on $z$, the fluctuation equations are diagonal in Fourier space, and it is sufficient to consider plane-wave modes of momentum $q$,
\begin{equation}
    \label{eq:planewavemodes}
    \zeta(x^\mu, z) = e^{iq \cdot x}\widehat{\zeta}(t,z) \ , \quad \varsigma(x^\mu,z) = e^{iq\cdot x}\widehat{\varsigma}(t,z) \ , \quad \hmntt(x^\mu,z) = e^{iq\cdot x}\varepsilon_\mn^{TT}(q)\widehat{h}(t,z) \ ,
\end{equation}
where $\varepsilon^{TT}_\mn(q)$ is a transverse-traceless polarization tensor, and the Fourier transformed radial profiles depend on the momentum only through $t = -q^2$. We define $\widehat{\F}(t,z)$ in the same way. The profiles $\widehat{\zeta}$ and $\widehat{\varsigma}$ solve a pair of coupled second-order equations first derived in~\cite{Arean:2013tja}, which we use in the form given in~\cite{Hippelainen:2026pion}, while $\widehat{h}$ solves the decoupled equation~\eqref{eq:tensorpropagator} below. Since the two scalars are coupled, there is a two-dimensional space of IR-regular solutions, which define the bulk-to-boundary propagators. However, the only scalar propagator we need is the one with equal boundary values of $\widehat{\zeta}$ and $\widehat{\varsigma}$, and for the tensor propagator we only need to fix the normalization. We choose
\begin{equation}
    \label{eq:propagatorboundaryvalues}
    \widehat{\zeta}(t,0) = \widehat{\varsigma}(t,0) = 1 \ , \qquad \widehat{h}(t,0) = 1 \ .
\end{equation}
That the two scalars have equal boundary values follows from the near-boundary behavior of the fluctuations, as we show in Appendix~\ref{app:boundaryvalues}.

The invariant transverse vector mode satisfies~\cite{Kiritsis:2006ua}
\begin{equation}
    \label{eq:vectoreoms}
   \Box \W^T_\mu = 0 \ , \qquad \qty(e^{3A}\W_\mu^T)' = 0 \ .
\end{equation}
In momentum space the first equation of~\eqref{eq:vectoreoms} reads $q^2 \widehat{\W}^T_\mu(q, z) = 0$, so $\widehat{\W}^T_\mu = 0$ at every $z$ for $q^2 \neq 0$, and the transverse vector carries no propagating degree of freedom. Once the meson equations of motion are used, the vertices depend on $A^T_\mu$ and $V^T_\mu$ only through $\W_\mu^T$ (Appendix~\ref{subapp:gaugeinvariance}).

\subsection{Vector and axial-vector mesons}
\label{subsec:vectorandaxialvectormesons}

The $\rho$ and $a_1$ are identified as the lowest modes of flavor non-singlet vector and transverse axial-vector fields~\cite{Arean:2013tja}, respectively. We derive their quadratic actions keeping the metric and the tachyon as general functions of $x^M$, since the same action is fluctuated in Section~\ref{sec:gravitationalformfactors} to obtain the cubic vertices. We write the tachyon as $T = \alphasymbol\qty(x^M) e^{i\Pi(x^M)}$, where $\alphasymbol$ is its modulus and $\Pi(x^M) = \pi^a(x^M) t^a$ its phase, with the pseudoscalar fields $\pi^a$, and the gauge fields in their vector and axial combinations,
\begin{equation}
    \label{eq:vectoraxialcombinations}
    V_M = \frac{L_M + R_M}{2} \ , \qquad A_M = \frac{L_M - R_M}{2} \ .
\end{equation}
Expanding the flavor action~\eqref{eq:vqcdaction} to quadratic order in $(\Pi, V_M, A_M)$ gives (Appendix~\ref{subapp:quadraticaction})
\begin{equation}
    \label{eq:quadraticflavoraction}
    \begin{split}
        S_f^{(2)} &= -\frac{1}{2}M^3 N_c \Tr \int \dd^5 x V_f\qty(\Phi, \alphasymbol^2)\sqrt{-\det\widetilde{g}} \\
        &\times \Big[2 + \kappa \alphasymbol^2 \widetilde{g}^\MN \qty(\pd_M \Pi + 2A_M)\qty(\pd_N \Pi + 2A_N) \\
        &\qquad + \frac{w^2}{2}F^V_\MN \widetilde{g}^{MO}\widetilde{g}^{NP}F^V_{OP} + \frac{w^2}{2}F^A_\MN \widetilde{g}^{MO}\widetilde{g}^{NP}F^A_{OP}\Big] \ ,
    \end{split}
\end{equation}
where $\widetilde{g}_\MN = g_\MN + \kappa (\pd_M \alphasymbol) (\pd_N \alphasymbol)$ is the effective metric~\eqref{eq:effectivemetric} for a general metric and tachyon modulus. The field strengths can be taken to be Abelian, since the commutator terms do not contribute at this order, so $F^V_\MN = \pd_M V_N - \pd_N V_M$, and analogously for $F^A_\MN$.

We split the axial field into its transverse and longitudinal parts, $A_\mu = A_\mu^\perp + A_\mu^\parallel$ with $\pd^\mu A^\perp_\mu = 0$ and $A^\parallel_\mu$ a gradient. On the background, the cross terms between $A^\perp_\mu$ and $(\Pi, A^\parallel_\mu)$ integrate to zero by transversality, so the transverse axial part decouples from the pion sector at quadratic order. The quadratic actions for vectors and transverse axial vectors then read
\begin{equation}
    \label{eq:quadraticactions}
    \begin{split}
        S_V^{(2)} &= -\frac{1}{4}M^3 N_c \Tr \int \dd^5 x \qty[b_1 V_\mn V^\mn + 2 b_2 V_{\mu z} V_{\nu z} \eta^\mn] \ , \\
        S_A^{(2)} &= -\frac{1}{4}M^3 N_c \Tr \int \dd^5 x \qty[b_1 A^\perp_\mn A^{\perp\mn} + 2 b_2 A^\perp_{\mu z} A_{\nu z}^\perp \eta^\mn + 8 c_1 A^\perp_\mu A^{\perp,\mu}] \ ,
    \end{split}
\end{equation}
where $V_\mn = \pd_\mu V_\nu - \pd_\nu V_\mu$, $V_{\mu z} = \pd_\mu V_z - \pd_z V_\mu$, and analogously for $A_\mu^\perp$. We have also defined the ubiquitous combinations
\begin{equation}
    \label{eq:b1b2c1}
    b_1 = V_f e^{A} G w^2 \ , \quad b_2 = \frac{b_1}{G^2} \ , \quad c_1 = V_f e^{3A}G \kappa \tau^2 \ , \quad c_2 = \frac{c_1}{G^2} \ ,
\end{equation}
which will appear in the vertices of Section~\ref{sec:gravitationalformfactors}. Compared to~\cite{Hippelainen:2026pion}, we have renamed the factors $a_i$ as $c_i$ to avoid a clash with the name of the axial meson. 

Equations of motion resulting from~\eqref{eq:quadraticactions} are
\begin{equation}
    \label{eq:eomsgeneral}
    \begin{split}
        b_1 \Box V_\mu + \pd_z \qty(b_2 \pd_z V_\mu) &= 0 \ , \\
        b_1 \Box A^\perp_\mu + \pd_z \qty(b_2 \pd_z A^\perp_\mu) - 4 c_1 A_\mu^\perp &= 0 \ .
    \end{split}
\end{equation}
Varying with respect to the radial components $V_z$ and $A_z$ gives constraints. We fix the radial gauge $V_z = A_z = 0$, which trivializes the vector constraint and makes the axial constraint affect only the pseudoscalar sector.

The actions~\eqref{eq:quadraticactions} and the equations of motion~\eqref{eq:eomsgeneral} of the vector and the transverse axial vector differ only by the term proportional to $c_1$, which is a $z$-dependent mass term induced by the tachyon background. We therefore treat both with a single field $B_\mu$, which stands for either $V_\mu$ or $A_\mu^\perp$, and write all expressions from now on for the general case, unless stated otherwise. The expressions for the vector are obtained by setting $c_1 \to 0$. Note that $V_\mu^T$ and $A_\mu^T$ are the transverse vector modes of the metric fluctuation~\eqref{eq:fluctuatedmetric}, while $V_\mu$ and $A_\mu^\perp$ are the flavor gauge fields.

We decompose the field into a radial profile and a four-dimensional mode of mass $m$,
\begin{equation}
    \label{eq:mesonmodedecomposition}
    B_\mu = \varphi(z) \B^a_\mu(x) t^a \ , \qquad \pd^\mu \B^a_\mu = 0 \ , \qquad \Box \B^a_\mu = m^2 \B^a_\mu \ ,
\end{equation}
so the equations~\eqref{eq:eomsgeneral} reduce to a single radial equation~\cite{Arean:2013tja},
\begin{equation}
    \label{eq:modeequation}
    \pd_z \qty(b_2 \pd_z \varphi) + m^2 b_1 \varphi - 4 c_1 \varphi = 0 \ .
\end{equation}
With regularity in the IR and normalizability in the UV, this is an eigenvalue problem with a discrete spectrum of masses in each channel. We identify the lowest vector state with $\rho$ and the lowest axial state with $a_1$, and denote by $m$ and $\varphi$ the mass and the profile of the state in question. At the order in $1/N_c$ at which we work, both states are stable.

\section{Gravitational form factors}
\label{sec:gravitationalformfactors}

We now compute the matrix element of the energy-momentum tensor between two states of the same spin-1 meson and identify six gravitational form factors from a decomposition of the matrix element. By the holographic dictionary, the boundary value of the metric fluctuation sources $T^\mn$ and the boundary value of the flavor gauge field sources the corresponding current, so the matrix element is contained in the part of the on-shell action that is linear in the metric fluctuation and quadratic in the meson field. Section~\ref{subsec:spin1matrixelementandlszreduction} defines the matrix element and the form factors. In Sections~\ref{subsec:dictionary} and~\ref{subsec:fieldtheorycorrelators} we set up the dictionary and express the matrix element through a three-point function of QCD. In Section~\ref{subsec:gravitycorrelators} we evaluate the two- and three-point functions on the gravity side, which fixes the normalization of the external states and results in the rule~\eqref{eq:vertexrule} for the matrix element. Sections~\ref{subsec:tensorvertex} and~\ref{subsec:scalarvertex} evaluate this rule for the tensor and scalar fluctuations. The derivation follows the one for the pion in~\cite{Hippelainen:2026pion}, where the LSZ reduction in the holographic setting is discussed in more detail, building on the holographic scattering formalism of~\cite{Hoyos:2019kzt, Hoyos:2020fjx, Hoyos:2022ptd}.

\subsection{Matrix element and form factors}
\label{subsec:spin1matrixelementandlszreduction}

We denote a state of the meson, $\rho$ or $a_1$, with flavor $a$, momentum $p$ and polarization $\epsilon$ by $\ket{B^a(p, \epsilon)}$, with $p^2 = -m^2$, $p \cdot \epsilon = 0$ and $\epsilon \cdot \epsilon^* = 1$. The states have relativistic normalization
\begin{equation}
    \label{eq:statenormalization}
    \braket{B^b(p', \epsilon')}{B^a(p, \epsilon)} = 2 E_{\boldsymbol{p}} (2\pi)^3 \delta^3(\boldsymbol{p} - \boldsymbol{p}') \delta^{ab} (\epsilon'^* \cdot \epsilon) \ ,
\end{equation}
where $E_{\boldsymbol{p}} = \sqrt{\boldsymbol{p}^2 + m^2}$, and the three polarization vectors $\epsilon(p, s)$ of a given momentum, labeled by $s$, satisfy
\begin{equation}
    \label{eq:polarizationsum}
    \sum_s \epsilon_\mu(p, s) \epsilon^*_\nu(p, s) = \eta_\mn + \frac{p_\mu p_\nu}{m^2} \ .
\end{equation}
For an incoming state with momentum $p_1$ and polarization $\epsilon_1$, and an outgoing state with $p_2$ and $\epsilon_2$, we use
\begin{equation}
    \label{eq:kinematics}
    q = p_2 - p_1 \ , \quad P = \frac{p_1 + p_2}{2} \ , \quad t = -q^2 \ , \quad \epsilon_{1q} = q \cdot \epsilon_1 \ , \quad \epsilon_{2q} = q \cdot \epsilon_2^* \ , \quad \epsilon_{12} = \epsilon_1 \cdot \epsilon_2^* \ ,
\end{equation}
and for on-shell states, $p_1^2 = p_2^2 = -m^2$, this implies $P \cdot q = 0$, $P^2 = -m^2 + t/4$, $p_1 \cdot p_2 = -m^2 + t/2$ and $t \leq 0$, while the transversality of the polarizations gives $P \cdot \epsilon_1 = \epsilon_{1q}/2$ and $P \cdot \epsilon_2^* = -\epsilon_{2q}/2$. We decompose the matrix element in the basis of~\cite{Polyakov:2019lbq}, which in mostly-plus signature reads
\begin{equation}
    \label{eq:decomposition}
    \begin{split}
        \mel{B^b(p_2, \epsilon_2)}{T^\mn(0)}{B^a(p_1, \epsilon_1)} = \delta^{ab}\Big\{
        & 2 P^\mu P^\nu \qty[\epsilon_{12} A_0(t) - \frac{\epsilon_{1q} \epsilon_{2q}}{4m^2} A_1(t)] \\
        & - \qty[P^{\mu}\qty(\epsilon_2^{*\nu} \epsilon_{1q} - \epsilon_1^{\nu}\epsilon_{2q}) + (\mu \leftrightarrow \nu)] J(t) \\
        & - \frac{1}{2}\qty(q^\mu q^\nu - q^2 \eta^\mn)\qty[\epsilon_{12} D_0(t) + \frac{\epsilon_{1q} \epsilon_{2q}}{4m^2} D_1(t)] \\
        & + \Big[\frac{\epsilon_{2q}}{2}\qty(\epsilon_1^\mu q^\nu + q^\mu \epsilon_1^\nu) + \frac{\epsilon_{1q}}{2}\qty(\epsilon_2^{*\mu} q^\nu + q^\mu \epsilon_2^{*\nu}) \\
        & - \frac{q^2}{2}\qty(\epsilon_1^\mu \epsilon_2^{*\nu} + \epsilon_2^{*\mu}\epsilon_1^{\nu}) - \eta^\mn \epsilon_{1q}\epsilon_{2q} \Big]E(t) \Big\} \ ,
    \end{split}
\end{equation}
where $T^\mn(0)$ is the operator at $x = 0$. The six form factors multiply the independent symmetric and conserved tensor structures that are bilinear in the two polarization vectors and even under parity and time reversal~\cite{Polyakov:2019lbq, Cosyn:2019aio}. The Poincar\'e transformation properties of the states fix $A_0(0) = J(0) = 1$~\cite{Lorce:2019sbq, Cotogno:2019xcl}, while the other four forward values depend on the dynamics. The quark and gluon parts of $T^\mn$ would separately have three further form factors multiplying non-conserved structures, but they cancel when quark and gluon contributions are summed into the total $T^\mn$~\cite{Polyakov:2019lbq}, the matrix element of which we compute. The relation of the form factors to spatial distributions and to the D-term is discussed in Section~\ref{subsec:breitframedensitiesandradii}.

\subsection{Setting up the dictionary}
\label{subsec:dictionary}

To obtain~\eqref{eq:decomposition} holographically, we start from the QCD generating functional. We perturb the metric of the field theory to $\eta_\mn + \delta\eta_\mn$ and couple a source $j^a_\mu$ to the vector or to the axial-vector current\footnote{The sign of the axial operator is chosen such that its source is the boundary value of the axial field $A_\mu$ of~\eqref{eq:vectoraxialcombinations}: with the left- and right-handed currents $J_{L,R}^{\mu a} = \bar{q}\gamma^\mu \frac{1}{2}(1 \mp \gamma_5) t^a q$ and their sources $l^a_\mu$ and $r^a_\mu$, one has $l^a_\mu J_L^{\mu a} + r^a_\mu J_R^{\mu a} = \frac{1}{2}(l + r)^a_\mu \bar{q}\gamma^\mu t^a q - \frac{1}{2}(l - r)^a_\mu \bar{q}\gamma^\mu\gamma_5 t^a q$. This sign does not affect the matrix elements we compute, in which $\O^{\mu a}$ appears twice.},
\begin{equation}
    \label{eq:qcdsources}
    \delta S_\text{QCD} = \int \dd^4 x \qty[\frac{1}{2}\delta\eta_\mn T^\mn + j^a_\mu \O^{\mu a}] \ , \qquad \O^{\mu a} = \bar{q}\gamma^\mu t^a q \quad \text{or} \quad \O^{\mu a} = -\bar{q}\gamma^\mu \gamma_5 t^a q \ .
\end{equation}
The generating functional $W[\delta\eta, j]$ of connected correlators is defined by $\mathcal{Z}[\delta\eta, j] = e^{iW[\delta\eta, j]}$, where $\mathcal{Z}$ is the QCD partition function in the presence of the sources. Since the metric couples to all fields of QCD, $T^\mn$ is the total energy-momentum tensor. The quark mass and gauge coupling are held fixed in the variation, so the sources of the tachyon and the dilaton remain unperturbed.

On the gravity side the sources are the boundary values of the corresponding bulk fields. As $z \to 0$,
\begin{equation}
    \label{eq:gravitysources}
    e^{-2A} g_\mn \to \eta_\mn + \delta \eta_\mn(x) \ , \qquad B^a_\mu(x, z) \to j^a_\mu(x) \ ,
\end{equation}
where $B_\mu = B^a_\mu t^a$ is the unified field, and the sources of the dilaton and of the tachyon are not perturbed. We take the source to be transverse, $\pd^\mu j^a_\mu = 0$, so in the axial case it sources only $A^\perp_\mu$, which decouples from the longitudinal part and from the pseudoscalar field at quadratic order (Section~\ref{subsec:vectorandaxialvectormesons}).

The two types of gravity fluctuations are sourced by different parts of $\delta\eta_\mn$. In momentum space, a symmetric tensor that is conserved with respect to the momentum $q$, such as the matrix element~\eqref{eq:decomposition} with $q_\mu T^\mn = 0$, decomposes into a transverse-traceless part and a part proportional to the transverse projector,
\begin{equation}
    \label{eq:TTtracesplit}
    \begin{split}
        &T^\mn = \varepsilon^{TT,\mn}{}_{\ab} T^\ab + \frac{1}{3}\Theta^\mn \Theta_\ab T^\ab \ , \qquad \Theta^\mn = \eta^\mn - \frac{q^\mu q^\nu}{q^2} \ , \\
        &\varepsilon^{TT,\mn}{}_{\ab} = \frac{1}{2}\qty(\Theta^\mu_{\ \alpha}\Theta^\nu_{\ \beta} + \Theta^\nu_{\ \alpha}\Theta^\mu_{\ \beta}) - \frac{1}{3}\Theta^\mn \Theta_\ab \ ,
    \end{split}
\end{equation}
which satisfy $q_\mu \Theta^\mn = 0$, $\Theta^\mu_{\ \mu} = 3$ and $q^\mu \TTproj = \eta^\mn \TTproj = 0$. By~\eqref{eq:boundaryvalues} of Appendix~\ref{app:boundaryvalues}, the transverse-traceless part of $\delta\eta_\mn$ sources the tensor fluctuation and its trace part the scalar fluctuations.

The dictionary then states that
\begin{equation}
    \label{eq:dictionary}
    W[\delta\eta, j] = S_\text{OS}[\delta\eta, j] \ ,
\end{equation}
where $S_\text{OS}$ is the on-shell gravity action, evaluated on the IR-regular solutions.

\subsection{Correlators in field theory}
\label{subsec:fieldtheorycorrelators}

We first express the matrix element~\eqref{eq:decomposition} through correlators that can be obtained from $W$. Throughout, the flavor index $a$ and the momentum $p_1$ refer to the incoming meson, and $b$ and $p_2$ to the outgoing one. The indices $a$ and $b$ are fixed and not summed over, while $c$ and $d$ denote generic flavor indices. The overlap of the operator with a one-meson state defines the decay constant $f$,
\begin{equation}
    \label{eq:decayconstant}
    \mel{0}{\O^{\mu c}(0)}{B^d(p, \epsilon)} = \delta^{cd} f m \epsilon^\mu \ , \qquad f > 0 \ .
\end{equation}
The connected two- and three-point functions are
\begin{equation}
    \label{eq:correlators}
    \begin{split}
        \delta^{cd} \widetilde{G}_2^{\mn}(p) &= i\int \dd^4 x e^{-ip\cdot x}\mel{0}{\mathsf{T}\qty{\O^{\mu c}(x) \O^{\nu d}(0)}}{0}_\text{conn} \ , \\
        \widetilde{G}_3^{\mn, \alpha\beta, ab}(p_2, p_1) &= -\int \dd^4 x_1 \dd^4 x_2 e^{ip_1\cdot x_1 - ip_2 \cdot x_2}\mel{0}{\mathsf{T}\qty{\O^{\beta b}(x_2) T^\mn(0) \O^{\alpha a}(x_1)}}{0}_\text{conn} \ ,
    \end{split}
\end{equation}
where $\mathsf{T}$ denotes time ordering, the subscript denotes the connected part, and $\alpha$ and $\beta$ are the Lorentz indices of the currents of the incoming and outgoing meson, respectively. Up to contact terms, $\widetilde{G}_2^{\mn}(p)$ and $\widetilde{G}_3^{\mn, \alpha\beta, ab}(p_2, p_1)$ are the Fourier transforms of $\delta^2 W/\delta j \delta j$ and $2\delta^3 W/\delta(\delta\eta)\delta j \delta j$. Inserting one-meson states and using~\eqref{eq:decayconstant}, the two-point function has a pole at $p^2 = -m^2$ with residue $f^2 m^2 \sum_s \epsilon^\mu(p,s) \epsilon^{*\nu}(p,s)$, and the three-point function has a double pole whose residue contains the matrix element. We contract the two current indices of $\widetilde{G}_3$ with polarization vectors $\epsilon_1$ and $\epsilon_2^*$ that are transverse to $p_1$ and $p_2$. By~\eqref{eq:polarizationsum} this selects the states with polarizations $\epsilon_1$ and $\epsilon_2$. The contraction also removes the contribution of the pion to the correlators of the axial current, since the overlap of the axial current with a pion of momentum $p$ is proportional to $p^\mu$. We obtain
\begin{equation}
    \label{eq:poles}
    \begin{split}
        \widetilde{G}_2^{\mn}(p) &= \frac{Z}{p^2 + m^2 - i0}\qty(\eta^\mn + \frac{p^\mu p^\nu}{m^2}) + \text{regular} \ , \qquad Z = f^2 m^2 \ , \\
        \epsilon_{1\alpha} \epsilon^*_{2\beta} \widetilde{G}_3^{\mn, \alpha\beta, ab}(p_2, p_1) &= \frac{Z \mel{B^b(p_2, \epsilon_2)}{T^\mn(0)}{B^a(p_1, \epsilon_1)}}{\qty(p_1^2 + m^2 - i0)\qty(p_2^2 + m^2 - i0)} + \text{terms without a double pole} \ .
    \end{split}
\end{equation}
LSZ reduction thus amounts to extracting the coefficient of the double pole and dividing by the $Z$ of the two external overlaps,
\begin{equation}
    \label{eq:LSZ}
    \mel{B^b(p_2, \epsilon_2)}{T^\mn(0)}{B^a(p_1, \epsilon_1)} = \frac{1}{Z}\lim_{p_i^2 \to -m^2}\qty(p_1^2 + m^2)\qty(p_2^2 + m^2) \epsilon_{1\alpha} \epsilon^*_{2\beta} \widetilde{G}_3^{\mn, \alpha\beta, ab}(p_2, p_1) \ .
\end{equation}

The three-point function is contained in the part of $W$ that is linear in $\delta\eta$ and quadratic in $j$,
\begin{equation}
    \label{eq:Wcubic}
    W^{(3)}[\delta \eta, j] = \frac{1}{4}\int \dd^4 x \dd^4 x_1 \dd^4 x_2 \delta\eta_\mn(x) j^c_\alpha(x_1) j^d_\beta(x_2) G_3^{\mn, \alpha\beta, cd}(x; x_1, x_2) \ ,
\end{equation}
where $G_3$ is the correlator of~\eqref{eq:correlators} in coordinate space with $T^\mn$ at the point $x$ and the currents at $x_1$ and $x_2$, and the factor $\frac{1}{4}$ is the product of the $\frac{1}{2}$ in~\eqref{eq:qcdsources} and the $1/2!$ of the two current sources. We choose the source
\begin{equation}
    \label{eq:sourcewaves}
    j^c_\mu(x) = \delta^{ca} \epsilon_{1\mu} e^{ip_1\cdot x} + \delta^{cb} \epsilon^*_{2\mu} e^{-ip_2\cdot x} \ .
\end{equation}
The index $c$ labels the components of the source and is summed over in~\eqref{eq:Wcubic}, while $a$ and $b$ are the fixed flavors of the incoming and outgoing meson. The momenta are off-shell, and polarization vectors are transverse to the momenta, $p_1 \cdot \epsilon_1 = p_2 \cdot \epsilon_2^* = 0$, so that the source is transverse. Let $W^{(3)}_{12}$ denote the part of $W^{(3)}$ that is bilinear in $\epsilon_1$ and $\epsilon_2^*$ for this source. The two ways of assigning the two plane waves to the two sources in~\eqref{eq:Wcubic} give equal contributions, since $G_3$ is symmetric under the simultaneous exchange of the labels of the two currents. By translation invariance we then obtain
\begin{equation}
    \label{eq:Wmixed}
    W^{(3)}_{12} = \frac{1}{2}\int \dd^4 x \delta\eta_\mn(x) e^{-iq\cdot x} \epsilon_{1\alpha} \epsilon^*_{2\beta} \widetilde{G}_3^{\mn, \alpha\beta, ab}(p_2, p_1) \ ,
\end{equation}
and therefore
\begin{equation}
    \label{eq:3pdictfinal}
    \epsilon_{1\alpha} \epsilon^*_{2\beta} \widetilde{G}_3^{\mn, \alpha\beta, ab}(p_2, p_1) = 2 e^{iq\cdot x} \frac{\delta W^{(3)}_{12}}{\delta\qty(\delta\eta_\mn(x))} \ .
\end{equation}

\subsection{Correlators in gravity}
\label{subsec:gravitycorrelators}

The field-theory analysis of Section~\ref{subsec:fieldtheorycorrelators} pins down what we need to compute in the holographic model. By the dictionary~\eqref{eq:dictionary}, the three-point function~\eqref{eq:3pdictfinal} is obtained from the cubic part of the on-shell gravity action, and the matrix element then follows from the limit~\eqref{eq:LSZ}. This limit also requires the residue $Z$ of the two-point function, which is the topic of Section~\ref{subsubsec:twopointfunction}. We then continue with the three-point function in Section~\ref{subsubsec:vertexrule}.

\subsubsection{The two-point function and the external states}
\label{subsubsec:twopointfunction}

We first fix the normalization of the external meson states in the holographic model. After the flavor trace, the quadratic action~\eqref{eq:quadraticactions} of the unified field $B_\mu = B_\mu^a t^a$ in the gauge $B_z = 0$ reads
\begin{equation}
    \label{eq:quadraticactionB}
    S^{(2)} = -\frac{M^3 N_c}{8}\sum_a \int \dd^4 x \dd z \qty[b_1 B^a_\mn B^{a\mn} + 2 b_2 \qty(\pd_z B^a_\mu) \qty(\pd_z B^{a\mu}) + 8 c_1 B^a_\mu B^{a\mu}] \ ,
\end{equation}
with $B^a_\mn = \pd_\mu B^a_\nu - \pd_\nu B^a_\mu$, and $b_1$, $b_2$ and $c_1$ defined in~\eqref{eq:b1b2c1}. A normalizable mode~\eqref{eq:mesonmodedecomposition} has a profile $\varphi$ that solves~\eqref{eq:modeequation}, is regular in the IR, and vanishes at the boundary as $\varphi \sim z^2$ up to logarithmic corrections~\cite{Arean:2013tja}. The equation fixes $\varphi$ only up to a constant factor, which we keep arbitrary. Inserting~\eqref{eq:mesonmodedecomposition} into~\eqref{eq:quadraticactionB} gives the action of a four-dimensional massive vector field,
\begin{equation}
    \label{eq:canonicalaction}
    S^{(2)} = -\int \dd^4 x \qty[\frac{\N}{4} \B^a_\mn \B^{a\mn} + \frac{m^2 \N}{2} \B^a_\mu \B^{a\mu}] \ , \qquad \N = \frac{M^3 N_c}{2}\int \dd z b_1 \varphi^2 \ .
\end{equation}
Here we used the fact that the coefficient of the mass term satisfies
\begin{equation}
    \label{eq:massidentity}
    \frac{M^3 N_c}{2}\int \dd z \qty[b_2 (\varphi')^2 + 4 c_1 \varphi^2] = m^2 \N \ ,
\end{equation}
which follows from multiplying~\eqref{eq:modeequation} by $\varphi$ and integrating by parts. The boundary term $b_2 \varphi \varphi'$ vanishes at both ends, in the UV since $b_2 \sim 1/z$ and $\varphi \sim z^2$ up to logarithms, and in the IR since $b_2$ vanishes there faster than any power of $z$ while $\varphi$ grows at most as a power. Hence $\sqrt{\N} \B^a_\mu$ is a canonically normalized field of mass $m$.

To relate the pole residue $Z$ to the normalization $\N$ of the radial profile, we compute the two-point function of $\O^{\mu a}$ holographically. This requires the solution of~\eqref{eq:eomsgeneral} in the presence of the source. For a transverse source, $p^\mu \widetilde{j}^a_\mu(p) = 0$ with $\widetilde{j}^a_\mu(p) = \int \dd^4 x e^{-ip\cdot x} j^a_\mu(x)$, the solution is given by a bulk-to-boundary propagator $K$,
\begin{equation}
    \label{eq:sourcedsolution}
    \begin{split}
        &B^a_\mu(x, z) = \int \frac{\dd^4 p}{(2\pi)^4} e^{ip\cdot x} K(p, z) \widetilde{j}^a_\mu(p) \ , \\
        &\pd_z\qty(b_2 \pd_z K) - \qty(p^2 b_1 + 4 c_1)K = 0 \ , \qquad K(p, 0) = 1 \ ,
    \end{split}
\end{equation}
where $K$ is regular in the IR. On this solution the bulk integrand of~\eqref{eq:quadraticactionB} is $B^a_\mu$ times its equation of motion up to a total radial derivative, and the boundary term in the IR vanishes for a regular solution. Only the term at the UV boundary remains,
\begin{equation}
    \label{eq:onshellaction}
    S^{(2)}_\text{OS} = \frac{M^3 N_c}{4}\sum_a \int \dd^4 x b_2 B^a_\mu \pd_z B^{a\mu}\Big|_{z \to 0} \ .
\end{equation}
Its variations with respect to the source give, up to local terms,
\begin{equation}
    \label{eq:G2holographic}
    \begin{split}
        \expval{\O^{\mu a}(x)} &= \frac{\delta W}{\delta j^a_\mu(x)} = \frac{M^3 N_c}{2}\lim_{z \to 0} b_2 \pd_z B^{a\mu}(x, z) \ , \\
        \widetilde{G}_2^{\mn}(p) &= \frac{M^3 N_c}{2}\lim_{z\to0} b_2 \pd_z K(p, z)\qty(\eta^\mn - \frac{p^\mu p^\nu}{p^2}) + \text{local terms} \ ,
    \end{split}
\end{equation}
where the projector appears because a transverse source probes only the transverse part of the two-point function. At $p^2 = -m^2$ it equals the polarization sum~\eqref{eq:polarizationsum}.

Next we need the behavior of $K$ near the pole. Let $u(p, z)$ be the IR-regular solution of the radial equation in~\eqref{eq:sourcedsolution} with a fixed normalization in the IR. At $p^2 = -m^2$ the radial equation is the mode equation~\eqref{eq:modeequation}, whose IR-regular solution is the normalizable mode, so $u(p, 0)$ vanishes. Thus, $K$ also has a first order pole,
\begin{equation}
    \label{eq:Kpole}
    K(p, z) = \frac{k \varphi(z)}{p^2 + m^2} + \text{regular at } p^2 = -m^2 \ ,
\end{equation}
with a constant $k$ that depends on the normalization of $\varphi$. Comparing~\eqref{eq:G2holographic} with~\eqref{eq:poles} gives
\begin{equation}
    \label{eq:Zflux}
    Z = f^2 m^2 = k \frac{M^3 N_c}{2}\lim_{z\to0} b_2 \varphi' \ .
\end{equation}
LSZ reduction needs $k$ in terms of $\N$, and the required relation follows from the radial equations. Multiplying the equation of $K$ in~\eqref{eq:sourcedsolution} by $\varphi$ and~\eqref{eq:modeequation} by $K$ and subtracting, the terms with $c_1$ cancel and
\begin{equation}
    \label{eq:wronskian}
    \dv{z}\qty[b_2\qty(K' \varphi - K \varphi')] = \qty(p^2 + m^2) b_1 K \varphi \ .
\end{equation}
We integrate this over $z$. The boundary term in the IR vanishes for two regular solutions, and at the UV boundary $K \to 1$ and $b_2 K' \varphi \to 0$, so that
\begin{equation}
    \label{eq:wronskianintegrated}
    \lim_{z\to0} b_2 \varphi' = \qty(p^2 + m^2) \int \dd z b_1 K \varphi \ .
\end{equation}
Taking the limit $p^2 \to -m^2$ with~\eqref{eq:Kpole} and multiplying by $M^3 N_c/2$, the right-hand side becomes $k$ times the integral $\N$ of~\eqref{eq:canonicalaction},
\begin{equation}
    \label{eq:fluxnorm}
    \frac{M^3 N_c}{2}\lim_{z\to0} b_2 \varphi' = k \N \ .
\end{equation}
Together with~\eqref{eq:Zflux} this gives $Z = k^2 \N$. Choosing the sign of $\varphi$ such that $k > 0$, we find
\begin{equation}
    \label{eq:residue}
    k = \frac{f m}{\sqrt{\N}} \ , \qquad f m = \frac{M^3 N_c}{2\sqrt{\N}}\lim_{z\to0} b_2 \varphi' \ .
\end{equation}
The second relation expresses the decay constant through the normalized mode, as in~\cite{Arean:2013tja}. With~\eqref{eq:Kpole}, the first relation implies
\begin{equation}
    \label{eq:legrule}
    \lim_{p^2 \to -m^2}\frac{p^2 + m^2}{\sqrt{Z}} K(p, z) = \frac{\varphi(z)}{\sqrt{\N}} \ .
\end{equation}
Comparing with the field-theory LSZ formula~\eqref{eq:LSZ},~\eqref{eq:legrule} can be understood as the amputation rule on the gravity side. In the on-shell limit, each bulk-to-boundary propagator $K$ of an external meson, multiplied by its factor $(p^2 + m^2)/\sqrt{Z}$, is replaced by the normalized profile $\varphi/\sqrt{\N}$.

\subsubsection{The three-point function and the vertex rule}
\label{subsubsec:vertexrule}

By the dictionary~\eqref{eq:dictionary}, $W^{(3)}$ of~\eqref{eq:Wcubic} equals the part of the on-shell action that is linear in the metric fluctuation and quadratic in the meson field, evaluated on the linear solutions sourced by $j$ and $\delta\eta_\mn$. This part is given by the cubic vertices that couple two fields $B_\mu$ to the gravity fluctuations, which we derive in Sections~\ref{subsec:tensorvertex} and~\ref{subsec:scalarvertex}. Corrections to the fields at second order in the sources have vanishing boundary values and satisfy the linearized equations up to source terms, so they do not contribute to the on-shell action at cubic order.

Following the field-theory computation of Section~\ref{subsec:fieldtheorycorrelators}, we now insert the source~\eqref{eq:sourcewaves}. Its Fourier transform is a sum of two delta functions, so the momentum integral in~\eqref{eq:sourcedsolution} can be carried out, and the meson field is
\begin{equation}
    \label{eq:sourcedlegs}
    B^c_\mu(x, z) = \delta^{ca} \epsilon_{1\mu} K(p_1, z) e^{ip_1\cdot x} + \delta^{cb} \epsilon^*_{2\mu} K(p_2, z) e^{-ip_2\cdot x} \ ,
\end{equation}
where we used that $K$ depends on the momentum only through $p^2$. The vertices are bilinear in $B_\mu$. We denote by $S^{(3)}_{12}$ the part of the cubic on-shell action that is bilinear in $\epsilon_1$ and $\epsilon_2^*$, so that $W^{(3)}_{12} = S^{(3)}_{12}$ with $W^{(3)}_{12}$ of~\eqref{eq:Wmixed}. By~\eqref{eq:3pdictfinal}, the holographic result for the three-point function is then
\begin{equation}
    \label{eq:3pdictholo}
    \epsilon_{1\alpha} \epsilon^*_{2\beta} \widetilde{G}_3^{\mn, \alpha\beta, ab}(p_2, p_1) = 2 e^{iq\cdot x} \frac{\delta S^{(3)}_{12}}{\delta\qty(\delta\eta_\mn(x))} \ ,
\end{equation}
where the meson fields in the vertices are the sourced solutions~\eqref{eq:sourcedlegs}, which contain the propagators $K$ rather than the normalizable modes. Each term of $S^{(3)}_{12}$ contains one factor of $K(p_1, z)$ or $\pd_z K(p_1, z)$ and one factor of $K(p_2, z)$ or $\pd_z K(p_2, z)$. These are multiplied by the plane wave $e^{-iq\cdot x}$, by the gravity fluctuations, and by polynomials in the momenta, which arise from the derivatives $\pd_\mu$ in the vertices.

We then apply the LSZ reduction~\eqref{eq:LSZ} to~\eqref{eq:3pdictholo}. By the structure of $S^{(3)}_{12}$ described above, the limit can be taken term by term. Since $Z = (f m)^2$, its prefactor factorizes over the two external states,
\begin{equation}
    \label{eq:LSZfactorized}
    \frac{\qty(p_1^2 + m^2)\qty(p_2^2 + m^2)}{Z} = \prod_{i = 1,2} \frac{p_i^2 + m^2}{\sqrt{Z}} \ ,
\end{equation}
and by~\eqref{eq:legrule} and its radial derivative, each factor replaces $K(p_i, z)$ by the normalized profile $\varphi(z)/\sqrt{\N}$, while the momenta in the polynomials take their on-shell values. More explicitly, we define the amputated action as
\begin{equation}
    \label{eq:Sampdef}
    \overline{S}^{(3)}_{12} = \lim_{p_i^2 \to -m^2} \frac{\qty(p_1^2 + m^2)\qty(p_2^2 + m^2)}{Z} S^{(3)}_{12} \ .
\end{equation}
Combining the above observations, the amputated action is obtained from $S^{(3)}_{12}$ by replacing the meson field~\eqref{eq:sourcedlegs} in the vertices with the normalizable mode~\eqref{eq:mesonmodedecomposition} and with
\begin{equation}
    \label{eq:amputatedlegs}
    \B^c_\mu(x) = \frac{1}{\sqrt{\N}}\qty(\delta^{ca} \epsilon_{1\mu} e^{ip_1\cdot x} + \delta^{cb} \epsilon^*_{2\mu} e^{-ip_2\cdot x}) \ ,
\end{equation}
where $p_1^2 = p_2^2 = -m^2$. The gravity fluctuations are not amputated and remain the solutions sourced by $\delta\eta_\mn$. 

In summary, combining~\eqref{eq:LSZ},~\eqref{eq:3pdictholo} and~\eqref{eq:Sampdef}, the vertex rule for the matrix element is
\begin{equation}
    \label{eq:vertexrule}
    \mel{B^b(p_2, \epsilon_2)}{T^\mn(0)}{B^a(p_1, \epsilon_1)} = 2 e^{iq\cdot x} \frac{\delta \overline{S}^{(3)}_{12}}{\delta\qty(\delta\eta_\mn(x))} \ ,
\end{equation}
where $\overline{S}^{(3)}_{12}$ is evaluated with the meson modes~\eqref{eq:amputatedlegs} and with the gravity fluctuations sourced by $\delta\eta_\mn$. The bilinears of the modes~\eqref{eq:amputatedlegs} that are bilinear in $\epsilon_1$ and $\epsilon_2^*$ are proportional to $e^{-iq\cdot x}$, so $\overline{S}^{(3)}_{12}$ depends on $\delta\eta_\mn$ only through its Fourier transform $\widetilde{\delta\eta}_\mn(q)$, and the right-hand side of~\eqref{eq:vertexrule} is independent of $x$. In the next two subsections we evaluate~\eqref{eq:vertexrule} for the tensor and the scalar fluctuations.

\subsection{Tensor vertex}
\label{subsec:tensorvertex}

The tensor and scalar fluctuations determine different parts of the matrix element. By the decomposition~\eqref{eq:TTtracesplit}, the tensor vertex determines the transverse-traceless part of~\eqref{eq:decomposition}, and the scalar vertex of Section~\ref{subsec:scalarvertex} its trace part.

Since $h^{TT}_\mn$ is diffeomorphism invariant on its own and does not mix with the scalars, the tensor vertex follows from~\eqref{eq:quadraticflavoraction} by the fluctuation $\widetilde{g}_\mn \mapsto \widetilde{g}_\mn + 2e^{2A} h^{TT}_\mn$, so that $\delta\widetilde{g}^\mn = -2 e^{-2A} h^{TT,\mn}$. The measure does not change, since $h^{TT}_\mn$ is traceless, and the total tensor fluctuation gives (Appendix~\ref{subapp:quadraticactionandcubicvertices})
\begin{equation}
    \label{eq:tensorvertex}
    \begin{split}
        &S^{(3)}_{hBB} = M^3 N_c \Tr \int \dd^5 x h^{TT}_\mn \qty[b_1 \sfY^\mn + b_2 \sfX^\mn + 4 c_1 \sfM^\mn] \ , \\
        &\sfX_\mn = \qty(\pd_z B_\mu) \qty(\pd_z B_\nu) \ , \qquad \sfY_\mn = B_{\mu\rho} B_\nu^{\ \rho} \ , \qquad \sfM_\mn = B_\mu B_\nu \ .
    \end{split}
\end{equation}
For $\rho$, $c_1 \to 0$ and the term with $\sfM_\mn$ is absent. By~\eqref{eq:sourcedfluctuations}, the Fourier transform of the tensor fluctuation is that of its boundary value multiplied by the propagator $\widehat{h}(t, z)$, which solves~\cite{DeWolfe:1999cp, Kiritsis:2006ua}
\begin{equation}
    \label{eq:tensorpropagator}
    \widehat{h}'' + 3A'\widehat{h}' + t \widehat{h} = 0 \ , \qquad \widehat{h}(t, 0) = 1 \ ,
\end{equation}
with regularity in the IR.

We then evaluate the tensor part of~\eqref{eq:vertexrule}. With $\Tr(t^a t^b) = \frac{1}{2}\delta^{ab}$, the parts bilinear in $\epsilon_1$ and $\epsilon_2^*$ are
\begin{equation}
    \label{eq:tensorbilinearsmodes}
    \begin{split}
    \Tr \sfY_\mn\big|_{12} &= \frac{\delta^{ab}}{2\N} \varphi^2 \qty[\epsilon_{12} p_{1\mu} p_{2\nu} - \epsilon_{1q} p_{1\mu}\epsilon^*_{2\nu} + \epsilon_{2q} \epsilon_{1\mu} p_{2\nu} + (p_1 \cdot p_2) \epsilon_{1\mu}\epsilon^*_{2\nu} + (\mu \leftrightarrow \nu)] e^{-iq\cdot x} \ , \\
        \Tr \sfX_\mn\big|_{12} &= \frac{\delta^{ab}}{2\N} (\varphi')^2 \epsilon_{1\{\mu}\epsilon^*_{2\nu\}} e^{-iq\cdot x} \ , \qquad
        \Tr \sfM_\mn\big|_{12} = \frac{\delta^{ab}}{2\N} \varphi^2 \epsilon_{1\{\mu}\epsilon^*_{2\nu\}} e^{-iq\cdot x} \ , 
    \end{split}
\end{equation}
where the braces denote symmetrization as in~\eqref{eq:fluctuatedmetric}. To compare TT-parts, we use the fact that two tensors have the same TT-part if they agree modulo $q^\mu$ and $\eta^\mn$, since $q^\mu \varepsilon_\mn^{TT, \ab} = \eta^\mn \varepsilon_\mn^{TT, \ab} = 0$. For the comparison we can therefore set $p_1 = p_2 = P$ and drop every term containing $q^\mu$ or $\eta^\mn$. In the decomposition~\eqref{eq:decomposition} this removes all $D$-terms and all but one $E$-term, and up to an overall factor of $\delta^{ab} \varepsilon_\mn^{TT,\ab}$, we are left with
\begin{equation}
\label{eq:physicalTT}
    2 P^\mu P^\nu \epsilon_{12} A_0 - \frac{P^\mu P^\nu \epsilon_{1q} \epsilon_{2q}}{2m^2}A_1 - \qty[P^\mu\qty(\epsilon_2^{*\nu}\epsilon_{1q} - \epsilon_1^\nu\epsilon_{2q}) + (\mu\leftrightarrow\nu)] J - \frac{q^2}{2}\epsilon_1^{\{ \mu}\epsilon_2^{* \nu \}}E \ .
\end{equation}
On the gravity side, the TT-part of the matrix element up to $\delta^{ab} \varepsilon_\mn^{TT,\ab}$ gives
\begin{equation}
    \label{eq:tensorTT}
        \T_\sfY \Big[P^\mu P^\nu \epsilon_{12} - P^\mu\qty(\epsilon_2^{*\nu}\epsilon_{1q} - \epsilon_1^\nu \epsilon_{2q}) \Big] + \qty[\qty(-m^2 + \frac{t}{2})\T_\sfY + \T_\sfX + \T_\sfM]\epsilon_1^{\mu}\epsilon_2^{*\nu} + (\mu \leftrightarrow \nu)\ ,
\end{equation}
where we have defined as shorthands the tensor integrals
\begin{equation}
    \label{eq:tensorintegrals}
    \begin{split}
        \T_\sfY(t) &= \frac{M^3 N_c}{2\N}\int \dd z \ b_1 \varphi^2 \widehat{h}(t, z) \ , \\
        \T_\sfX(t) &= \frac{M^3 N_c}{2\N}\int \dd z \ b_2 (\varphi')^2 \widehat{h}(t, z) \ ,\\
        \T_\sfM(t) &= \frac{2 M^3 N_c}{\N}\int \dd z \ c_1 \varphi^2 \widehat{h}(t, z) \ .
    \end{split}
\end{equation}
Directly comparing~\eqref{eq:physicalTT} with~\eqref{eq:tensorTT}, we identify the four tensor form factors
\begin{equation}
    \label{eq:tensorFFs}
    A_0(t) = J(t) = \T_\sfY(t) \ , \qquad A_1(t) = 0 \ , \qquad E(t) = \T_\sfY(t) - \frac{2}{t}\qty[m^2 \T_\sfY(t) - \T_\sfX(t) - \T_\sfM(t)] \ .
\end{equation}

In this model, the relations $A_0 = J$ and $A_1 = 0$ hold identically for both mesons, since the field-strength bilinear $\sfY_\mn$ produces the $A_0$ and $J$ structures with the same factor, and since no $P^\mu P^\nu \epsilon_{1q}\epsilon_{2q}$ structure appears.

\subsection{Scalar vertex}
\label{subsec:scalarvertex}

The scalar vertex is the first-order fluctuation of~\eqref{eq:quadraticflavoraction} with respect to the dilaton, the tachyon and the scalar and transverse vector parts of the metric~\eqref{eq:fluctuatedmetric}. We derive it in detail in Appendix~\ref{subapp:quadraticactionandcubicvertices}. First, the coefficients of the gauge modes $\chi$, $E$ and $V^T_\mu$ vanish once the equations of motion~\eqref{eq:eomsgeneral} and the background equations are used, which confirms that the vertex is gauge invariant (Appendix~\ref{subapp:gaugeinvariance}). Second, we set the gauge modes to zero and write the remaining fluctuations in terms of the invariants with~\eqref{eq:invertedscalars}. Third, we bring the vertex to a canonical form. Up to vanishing total derivatives, the gauge invariant scalar vertex can be written as
\begin{equation}
    \label{eq:scalarvertex}
    S^{(3)}_{sBB} = -\frac{M^3 N_c}{2}\Tr \int \dd^5 x \qty[\sum_{I \in \qty{\sfX, \sfY, \sfM}} I\qty(c_{I\zeta} \zeta + c_{I\varsigma} \varsigma + c_{I\F} \F) - 2 b_2 \W^{T,\mu} \sfZ_\mu] \ ,
\end{equation}
where we have defined one new bilinear, $\sfZ_\mu = B_\mn (B')^\nu$, and $\sfX = \eta^\mn \sfX_\mn$ are the traces of the bilinears (analogously for others). By~\eqref{eq:vectoreoms}, $\widehat{\W}^T_\mu(q, z) = 0$ for $q^2 \neq 0$, so the last term of~\eqref{eq:scalarvertex} does not contribute to the matrix element. With $b_1$, $b_2$, $c_1$ and $c_2$ as in~\eqref{eq:b1b2c1} and $G$ as in~\eqref{eq:effectivemetric}, the coefficient functions are
\begin{equation}
    \label{eq:scalarcoefficients}
    \begin{split}
        c_{\sfY\zeta} &= \frac{b_1}{2}\qty[\sigma + \frac{\tau' \pd_\tau V_f}{A' V_f} + \frac{4\qty(\Phi')^2}{9 G^2 (A')^2}] - \qty(\frac{b_1}{2A'})' \ , \\
        c_{\sfY\varsigma} &= -\frac{b_1}{2}\qty[\sigma + \frac{\tau' \pd_\tau V_f}{A' V_f} - \frac{x V_f \kappa \qty(\tau')^2}{6 G^3 (A')^2}] + \qty(\frac{b_1 - b_2}{2A'})' \ , \\
        c_{\sfY\F} &= \frac{1}{2}\qty(b_2' - 3A' b_2) \ , \\
        c_{\sfX\zeta} &= b_2\qty[2 - \sigma + \frac{\tau' \pd_\tau V_f}{A' V_f} - \frac{4\qty(\Phi')^2}{9 G^2 (A')^2}] + \qty(\frac{b_2}{A'})' - \frac{2 b_2'}{A'} \ , \\
        c_{\sfX\varsigma} &= b_2\qty[\sigma - \frac{\tau' \pd_\tau V_f}{A' V_f} - \frac{x V_f \kappa \qty(\tau')^2}{6 G^3 (A')^2}] - \qty(\frac{(G^2 - 1) b_2}{G^2 A'})' + \frac{2 (G^2 - 1) b_2'}{G^2 A'} \ , \\
        c_{\sfX\F} &= \frac{2 b_2'}{G^2} - \qty(\frac{b_2}{G^2})' + \frac{3 A' b_2}{G^2} \ , \\
        c_{\sfM\zeta} &= 4 c_1\qty[2 + \sigma + \frac{\tau'}{A'}\qty(\frac{\pd_\tau V_f}{V_f} + \frac{2}{\tau}) + \frac{4\qty(\Phi')^2}{9 G^2 (A')^2}] - \qty(\frac{4 c_1}{A'})' \ , \\
        c_{\sfM\varsigma} &= -4 c_1\qty[\sigma + \frac{\tau'}{A'}\qty(\frac{\pd_\tau V_f}{V_f} + \frac{2}{\tau}) - \frac{x V_f \kappa \qty(\tau')^2}{6 G^3 (A')^2}] + \qty(\frac{4 (G^2 - 1) c_1}{G^2 A'})' \ , \\
        c_{\sfM\F} &= 4\qty(c_2' - 3A' c_2) \ ,
    \end{split}
\end{equation}
where
\begin{equation}
    \label{eq:sigmavtau}
    \sigma = \frac{1}{G^2}\qty[\frac{x V_f \kappa \qty(\tau')^2}{6 G (A')^2} + e^{-2A}\kappa \tau' \qty(\frac{\tau'}{A'})'] \ .
\end{equation}
To identify the remaining form factors, we consider the trace part $\frac{1}{3}\Theta^\mn \Theta_\ab T^\ab$ of the decomposition~\eqref{eq:TTtracesplit}. By~\eqref{eq:sourcedfluctuations}, the Fourier transforms of the three scalar invariants are their propagators multiplied by the same boundary factor $\widetilde{\psi}^{(0)}(q) = \frac{1}{6}\Theta^\mn(q) \widetilde{\delta\eta}_\mn(q)$, so the propagators enter the vertex~\eqref{eq:scalarvertex} only through the combinations
\begin{equation}
    \label{eq:scriptC}
    \C_I(t, z) = c_{I\zeta} \widehat{\zeta}(t, z) + c_{I\varsigma} \widehat{\varsigma}(t, z) + c_{I\F} \widehat{\F}(t, z) \ , \qquad I \in \qty{\sfX, \sfY, \sfM} \ .
\end{equation}
The parts of the traced factors that are bilinear in $\epsilon_1$ and $\epsilon_2^*$ are
\begin{equation}
    \label{eq:scalarbilinearsmodes}
    \begin{split}
    &\Tr \sfY\big|_{12} = \frac{2\delta^{ab}}{\N}\varphi^2 \qty[(p_1 \cdot p_2) \epsilon_{12} + \epsilon_{1q}\epsilon_{2q}] e^{-iq\cdot x} \ , \\
        &\Tr \sfX\big|_{12} = \frac{\delta^{ab}}{\N}(\varphi')^2 \epsilon_{12} e^{-iq\cdot x} \ , \\
        &\Tr \sfM\big|_{12} = \frac{\delta^{ab}}{\N}\varphi^2 \epsilon_{12} e^{-iq\cdot x} \ .
    \end{split}
\end{equation}
Since the boundary factor depends on the source only through $\frac{1}{6}\Theta^\mn(q) \widetilde{\delta\eta}_\mn(q)$, the vertex rule~\eqref{eq:vertexrule} gives a contribution to the matrix element proportional to $\Theta^\mn$, whose trace, up to the factor $\delta^{ab}$, is
\begin{equation}
    \label{eq:scalaramplitude}
    \Theta_\mn \mel{B^b(p_2, \epsilon_2)}{T^\mn(0)}{B^a(p_1, \epsilon_1)} = -\frac{1}{2}\Big\{\qty[\S_\sfX + \S_\sfM - \qty(m^2 - \frac{t}{2})\S_\sfY]\epsilon_{12} + \S_\sfY \epsilon_{1q} \epsilon_{2q}\Big\} \ ,
\end{equation}
where we have defined scalar integrals analogous to the ones in the tensor computation,
\begin{equation}
    \label{eq:scalarintegrals}
    \begin{split}
        \S_\sfX(t) &= \frac{M^3 N_c}{\N}\int \dd z \ \C_\sfX (\varphi')^2 \ , \\
        \S_\sfY(t) &= \frac{2M^3 N_c}{\N}\int \dd z \ \C_\sfY \varphi^2 \ , \\
        \S_\sfM(t) &= \frac{M^3 N_c}{\N}\int \dd z \ \C_\sfM \varphi^2 \ ,
    \end{split}
\end{equation}
and where we used~\eqref{eq:scriptC}. Similarly, contracting~\eqref{eq:decomposition} with $\Theta_\mn$ gives, up to the factor $\delta^{ab}$,
\begin{equation}
    \label{eq:physicaltrace}
    \begin{split}
        \Theta_\mn &\mel{B^b(p_2, \epsilon_2)}{T^\mn(0)}{B^a(p_1, \epsilon_1)} \\
        &= \epsilon_{12}\qty[2P^2 A_0 - \frac{3t}{2}D_0 + t E] + \epsilon_{1q}\epsilon_{2q}\qty[-\frac{P^2}{2m^2}A_1 + 2J - \frac{3t}{8m^2}D_1 - 2E] \ .
    \end{split}
\end{equation}
With $P^2 = -m^2 + t/4$ and with $A_0$, $A_1$, $J$ and $E$ fixed by the tensor channel~\eqref{eq:tensorFFs}, the coefficients of $\epsilon_{12}$ and $\epsilon_{1q}\epsilon_{2q}$ in~\eqref{eq:scalaramplitude} and~\eqref{eq:physicaltrace} determine the remaining two form factors in terms of the tensor and scalar integrals,
\begin{equation}
    \label{eq:D0D1}
    \begin{split}
        D_0(t) &= \frac{1}{3t}\qty[\S_\sfX + \S_\sfM - \qty(m^2 - \frac{t}{2})\S_\sfY - \qty(8m^2 - 3t)\T_\sfY + 4\qty(\T_\sfX + \T_\sfM)] \ , \\
        D_1(t) &= \frac{8m^2}{3t}\qty[\frac{1}{2}\S_\sfY + \frac{4}{t}\qty(m^2 \T_\sfY - \T_\sfX - \T_\sfM)] \ .
    \end{split}
\end{equation}
For $\rho$, $\T_\sfM = \S_\sfM = 0$. The expressions~\eqref{eq:tensorFFs} and~\eqref{eq:D0D1} contain explicit factors of $1/t$ in $E$, $D_0$ and $D_1$. We show in Appendix~\ref{subapp:regularity} that the factors multiplying the apparent poles vanish at $t = 0$ analytically, and that all form factors are finite.

\section{Results}
\label{sec:results}

We now evaluate the form factors of Section~\ref{sec:gravitationalformfactors} numerically, with the methods of~\cite{Hippelainen:2026pion}. The mode equation~\eqref{eq:modeequation} is solved as an eigenvalue problem, the propagators $\widehat{h}$, $\widehat{\zeta}$ and $\widehat{\varsigma}$ as boundary-value problems in $z$, and $\widehat{\F}$ is obtained by integrating~\eqref{eq:conjugateequation}. The mass of the lowest vector state is $m_\rho = 775.3$~MeV by the choice of the energy scale in Section~\ref{subsec:vqcdandthebackground}, and the lowest axial state has $m_{a_1} = 1233$~MeV.

Although the poles in~\eqref{eq:tensorFFs} and~\eqref{eq:D0D1} cancel exactly, numerically $E$, $D_0$ and $D_1$ are small differences of terms that each grow as inverse powers of $t$, so their relative error grows as $t \to 0$. We therefore obtain the forward values of $E$, $D_0$ and $D_1$ from quadratic fits in the lowest non-zero values of $t$ on the numerical grid. For $E(0)$ the fit agrees with the closed form~\eqref{eq:E0closedform}.

\subsection{Form factors of the \texorpdfstring{$\rho$ and $a_1$}{rho and a1}}
\label{subsec:formfactorsoftherhoanda1}

\begin{figure}[htb]
    \centering
    \includegraphics[width=\linewidth]{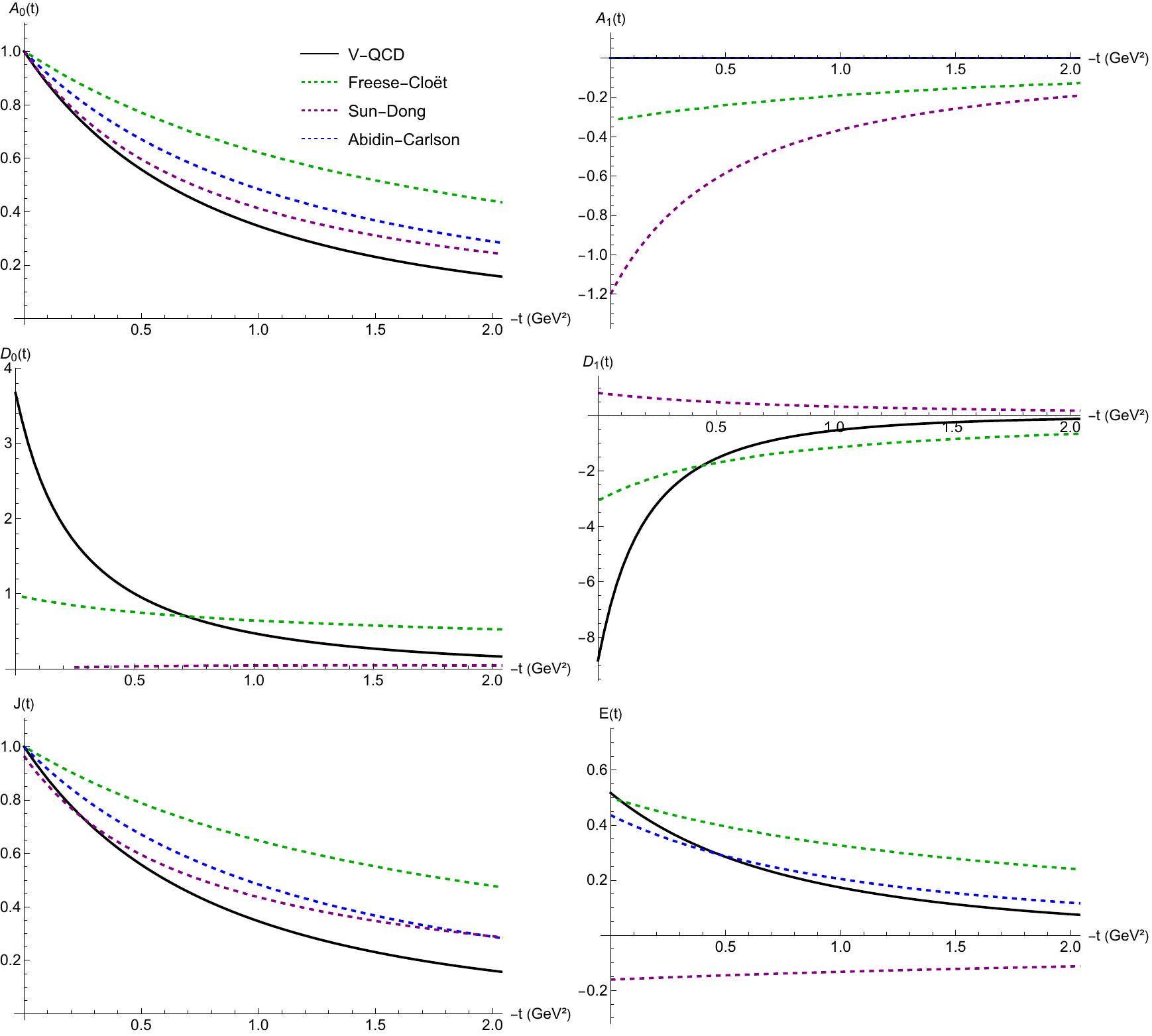}
    \caption{The six gravitational form factors of the $\rho$ meson in the basis~\eqref{eq:decomposition}: V-QCD (solid black), the NJL model result of~\cite{Freese:2019bhb} (green dashed), the light-cone constituent quark model result of~\cite{Sun:2020wfo} (purple dashed, from their closed-form fits) and the hard-wall result of~\cite{Abidin:2008ku} (blue dashed). Both holographic models have $A_0(t) = J(t)$ and $A_1(t) = 0$, and~\cite{Abidin:2008ku} does not determine $D_0(t)$ and $D_1(t)$. The results of the other works were converted to the basis~\eqref{eq:decomposition} with the dictionary of~\cite{Polyakov:2019lbq}.}
    \label{fig:rhoGFFs}
\end{figure}

Figure~\ref{fig:rhoGFFs} shows the six form factors of $\rho$ together with three other determinations of the total form factors: the NJL model-based determination of Freese and Clo\"et~\cite{Freese:2019bhb}, the light-cone constituent quark model determination of Sun and Dong~\cite{Sun:2020wfo}, and the hard-wall holographic results of Abidin and Carlson~\cite{Abidin:2008ku}. All are shown in the basis~\eqref{eq:decomposition}, using the dictionary of~\cite{Polyakov:2019lbq}. The only lattice results for the $t$ dependence of the form factors of $\rho$ that we are aware of are for the gluon contributions at a pion mass $m_\pi = 450$~MeV~\cite{Pefkou:2021fni}, so we cannot reliably compare with them.

A few notes are in order: First, $A_0 = J$ and $A_1 = 0$ hold identically, in V-QCD as in the hard-wall model~\cite{Abidin:2008ku}. The NJL and light-cone results have $A_1 \neq 0$ and $J \neq A_0$. Second, $E(0)$ differs from its free-field value of one~\cite{Polyakov:2019lbq}: we find $E(0) = 0.517$ for $\rho$.

In the range shown in Figure~\ref{fig:rhoGFFs}, $A_0$ and $J$ decrease faster with increasing $-t$ in V-QCD than in the other three approaches, and the $E$ in V-QCD decreases faster than the $E$ of the NJL and hard-wall models. The differences are largest in $D_0$ and $D_1$, which are determined by the scalar channel. In V-QCD they are larger in magnitude near $t = 0$ than in the NJL model and approach zero faster, while the light-cone fits have $D_0 \approx 0$ and the opposite sign for $D_1$. The light-cone result for $E$ is also negative, while the other three are positive.

\begin{figure}[htb]
    \centering
    \includegraphics[width=0.75\linewidth]{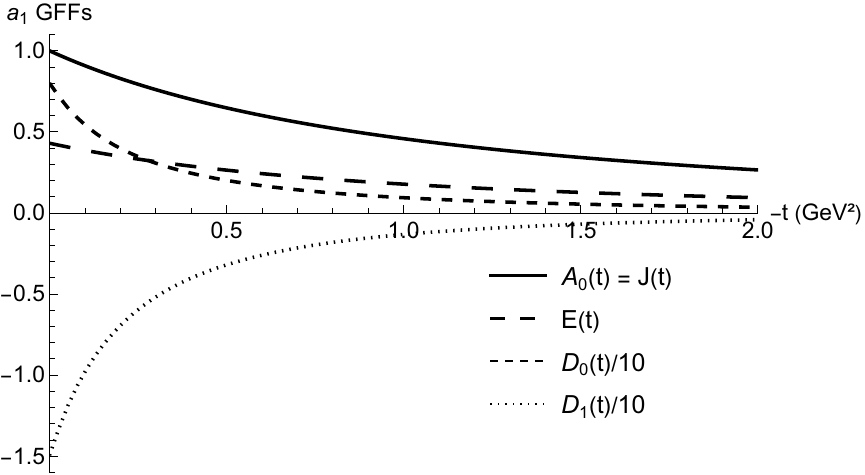}
    \caption{The gravitational form factors of the $a_1$ meson in V-QCD. As for $\rho$, $A_0 = J$ and $A_1 = 0$ identically. The form factors $D_0$ and $D_1$ have been divided by $10$ for readability.}
    \label{fig:a1GFFs}
\end{figure}

Figure~\ref{fig:a1GFFs} shows the form factors of $a_1$. Compared with $\rho$, $A_0$ falls off more slowly, $E(0) = 0.430$ is lower, and $D_0$ and $D_1$ are about twice as large at small $-t$, with $D_0(0) = 8.01$ and $D_1(0) = -14.9$ against $3.67$ and $-8.84$ for $\rho$.

\subsection{Breit-frame form factors, densities and radii}
\label{subsec:breitframedensitiesandradii}

For spin-0 and spin-1/2 hadrons, pressure and shear force distributions can be defined from the single form factor $D(t)$, whose forward value is the D-term~\cite{Polyakov:2018zvc}. A polarized spin-1 state is not spherically symmetric, and its energy density and stress tensor have both monopole and quadrupole parts~\cite{Polyakov:2019lbq}. The construction has two steps: First, in momentum space, the matrix element~\eqref{eq:decomposition} is evaluated in the Breit frame, where it is organized by six combinations of the form factors. We call them the Breit-frame form factors, and they are functions of $t$. Second, in position space, the Fourier transform of the Breit-frame matrix element defines the static energy-momentum tensor, whose multipole factors are the spatial distributions of energy, pressure and shear force. They are functions of the distance $r$ from the center of the meson, and each of them is obtained from one of the Breit-frame form factors.

\begin{figure}[htb]
    \centering
    \includegraphics[width=\linewidth]{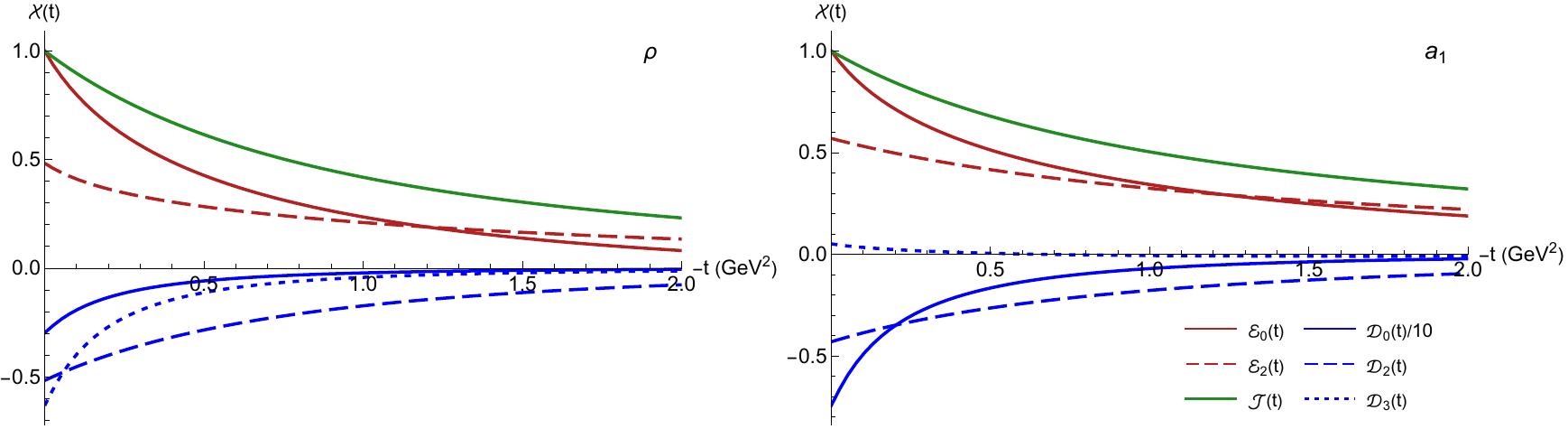}
    \caption{The Breit-frame form factors~\eqref{eq:breitcombinations} for $\rho$ (left) and $a_1$ (right): energy monopole $\E_0$ and quadrupole $\E_2$, spin $\J$, and the form factors $\D_0$, $\D_2$ and $\D_3$ of the stress tensor. Monopoles are shown with solid lines, quadrupoles with dashed lines. The legend on the right applies to both plots.}
    \label{fig:breitcombinations}
\end{figure}

In the Breit frame, $P^\mu = (E_q, \vb{0})$ and $q^\mu = (0, \vb{q})$, so that $t = -\vb{q}^2$, and $E_q = \sqrt{m^2 + \vb{q}^2/4}$ is the energy of both the incoming and the outgoing meson. In this frame the matrix element of the total EMT takes the form~\cite{Polyakov:2019lbq}
\begin{equation}
    \label{eq:breitmatrixelement}
    \begin{split}
        \mel{B(p_2)}{T^{00}(0)}{B(p_1)} &= 2m^2 \E_0(t) + Q^{kl} q^k q^l \E_2(t) \ , \\
        \mel{B(p_2)}{T^{0j}(0)}{B(p_1)} &= i \epsilon^{jkl} S^k q^l m \J(t) \ , \\
        \mel{B(p_2)}{T^{ij}(0)}{B(p_1)} &= \frac{1}{2}\qty(q^i q^j - \delta^{ij}\vb{q}^2)\D_0(t) \\
        &+ \qty(q^j q^k Q^{ik} + q^i q^k Q^{jk} - \vb{q}^2 Q^{ij} - \delta^{ij} q^k q^l Q^{kl})\D_2(t) \\
        &+ \frac{1}{2m^2}\qty(q^i q^j - \delta^{ij}\vb{q}^2) q^k q^l Q^{kl} \D_3(t) \ ,
    \end{split}
\end{equation}
where the flavor factor $\delta^{ab}$ and the polarization labels of the states are suppressed,  $S^k$ is the spin operator and $Q^{kl} = \frac{1}{2}\qty(S^k S^l + S^l S^k) - \frac{2}{3}\delta^{kl}$ is the quadrupole operator of a spin-1 state. The Latin indices $i, j, k, l = 1, 2, 3$ are spatial. The terms without $Q^{kl}$ are the monopole parts and those with $Q^{kl}$ the quadrupole parts. For the total EMT, and with $A_1 = 0$ and $J = A_0$, the six Breit-frame form factors of eq.~(17) of~\cite{Polyakov:2019lbq} are defined explicitly as
\begin{equation}
    \label{eq:breitcombinations}
    \begin{split}
        \E_0 &= A_0 + \frac{t}{12m^2}\qty(-A_0 + 3D_0 - 2E) - \frac{t^2}{24m^4}\qty(A_0 + D_0 - 2E + \frac{1}{2}D_1) + \frac{t^3}{192m^6}D_1 \ , \\
        \E_2 &= A_0 - E - \frac{t}{4m^2}\qty(A_0 + D_0 - 2E + \frac{1}{2}D_1) + \frac{t^2}{32m^4}D_1 \ , \\
        \J &= A_0 - \frac{t}{4m^2}\qty(A_0 - E) \ , \\
        \D_0 &= -D_0 + \frac{4}{3}E + \frac{t}{12m^2}\qty(2D_0 - 2E + D_1) - \frac{t^2}{48m^4}D_1 \ , \\
        \D_2 &= -E \ , \\
        \D_3 &= \frac{1}{4}\qty(2D_0 - 2E + D_1) - \frac{t}{16m^2}D_1 \ .
    \end{split}
\end{equation}
Figure~\ref{fig:breitcombinations} shows these six Breit-frame form factors for both mesons. In the range shown, $\E_0$, $\E_2$ and $\J$ are positive and decrease monotonically with $-t$, while all three $\D_n$ are negative for $\rho$. For $a_1$, $\D_3$ instead comes out small and positive near $t = 0$.

Since for a spin-1 hadron the stress tensor is not given by a single form factor, the D-term is defined in~\cite{Polyakov:2019lbq} as the forward value of its monopole Breit frame form factor $\D_0$, $\D_0(0) = -D_0(0) + \frac{4}{3}E(0)$. We find
\begin{equation}
    \label{eq:dterms}
    \D_0^\rho(0) \approx -2.98 \ , \qquad \D_0^{a_1}(0) \approx -7.43 \ ,
\end{equation}
both of which are dominated by $-D_0(0)$. For $\rho$, $\D_0(0) \approx -0.21$ in the light-cone constituent quark model of~\cite{Sun:2020wfo}, where also the value $\D_0(0) \approx -0.33$ is quoted for the profiles of~\cite{Freese:2019bhb}. Our prediction is thus an order of magnitude larger in magnitude. We note that other definitions are also in use: for example, in~\cite{Stegeman:2025tdl} the D-term of $\rho$ is defined as $-D_0(0)$, for which we find $-3.67$.

The static EMT is defined as the Fourier transform of the Breit-frame matrix element~\eqref{eq:breitmatrixelement} with respect to $\vb{q}$~\cite{Polyakov:2018zvc, Polyakov:2019lbq},
\begin{equation}
    \label{eq:staticEMT}
    T^\mn(\vb{r}) = \int \frac{\dd^3 q}{(2\pi)^3 2E_q} e^{-i\vb{q}\cdot\vb{r}} \mel{B(p_2)}{T^\mn(0)}{B(p_1)} \ .
\end{equation}
Under the transform, the energy density and the stress tensor take the form of a multipole expansion~\cite{Polyakov:2019lbq},
\begin{equation}
    \label{eq:breitdecomposition}
    \begin{split}
        T^{00}(\vb{r}) &= \varepsilon_0(r) + \varepsilon_2(r) Q^{kl} Y_2^{kl} \ , \\ 
        T^{ij}(\vb{r}) &= p_0(r) \delta^{ij} + s_0(r) Y_2^{ij} + p_2(r) Q^{ij} +2 s_2(r) \qty[Q^{ik} Y_2^{kj} + Q^{jk}Y_2^{ki} - \delta^\ij Q^{kl}Y_2^{kl}] \\
        &- \frac{1}{m^2} Q^{kl} \pd_k \pd_l \qty[p_3(r) \delta^\ij + s_3(r) Y_2^\ij] \ ,
    \end{split}
\end{equation}
where $Y_2^{ij} = \widehat{r}^i \widehat{r}^j - \frac{1}{3}\delta^{ij}$ is an irreducible tensor of rank two and $r = |\vb{r}|$. The radial functions $\varepsilon_0$ and $\varepsilon_2$ are the monopole and quadrupole distributions of energy, and $p_n$ and $s_n$ with $n = 0, 2, 3$ those of pressure and shear force. Each distribution is determined by a Breit-frame form factor $X(t)$ through a transform
\begin{equation}
    \label{eq:breittransform}
    \widetilde{X}(r) = \int \frac{\dd^3 q}{(2\pi)^3 2E_q} e^{-i\vb{q}\cdot\vb{r}} X(-q^2) \ ,
\end{equation}
where $q = |\vb{q}|$. The distributions are then~\cite{Polyakov:2019lbq}
\begin{equation}
    \label{eq:densities}
    \varepsilon_0 = 2m^2 \widetilde{\E}_0 \ , \quad \varepsilon_2 = -r\dv{r}\frac{1}{r}\dv{r}\widetilde{\E}_2 \ , \quad p_n = \frac{1}{3r^2}\dv{r} r^2\dv{r}\widetilde{\D}_n \ , \quad s_n = -\frac{1}{2}r\dv{r}\frac{1}{r}\dv{r}\widetilde{\D}_n \ .
\end{equation}
After the angular integration, the $r^2$-weighted distributions that we show below are obtained from the one-dimensional integrals
\begin{equation}
    \label{eq:densityintegrals}
    \begin{split}
        r^2 \varepsilon_0 &= \frac{m^2 r}{2\pi^2}\int_0^\infty \dd q \frac{q}{E_q} \E_0(-q^2) \sin(qr) \ , \qquad r^2 \varepsilon_2 = -\frac{r^2}{4\pi^2}\int_0^\infty \dd q \frac{q^4}{E_q} \E_2(-q^2) j_2(qr) \ , \\
        r^2 p_n &= -\frac{r}{12\pi^2}\int_0^\infty \dd q \frac{q^3}{E_q} \D_n(-q^2) \sin(qr) \ , \qquad r^2 s_n = -\frac{r^2}{8\pi^2}\int_0^\infty \dd q \frac{q^4}{E_q} \D_n(-q^2) j_2(qr) \ ,
    \end{split}
\end{equation}
where $j_2$ is the spherical Bessel function of order two.

These integrals require the form factors at all momentum transfers, but since it is hard to extend the holographic computation to asymptotically large momentum transfers, we continue the numerical form factors beyond $-t = 6.4$~GeV$^2$ with power-law tails, as for the pion~\cite{Hippelainen:2026pion}. Each tail is fitted to the numerical results in the window $4 \leq -t \leq 6.4$~GeV$^2$, and consists of the leading powers $A_0, E, D_0 \sim 1/t^2$ and $D_1 \sim 1/t^3$~\cite{Abidin:2008ku,Sun:2020wfo}, and the next three subleading integer powers.

The leading powers matter more than for spin-0 hadrons, since in~\eqref{eq:breitcombinations} the form factors are multiplied by powers of $t/m^2$ up to $(t/m^2)^3$. With the leading powers above, $\E_0$ tends to a constant and $\D_0$ falls off as $1/t$ at large $-t$, and so $r^2 \varepsilon_0$, $r^2 p_0$ and $r^2 s_0$ tend to finite constants as $r \to 0$. One power of $t$ less in any of the form factors would make $\E_0$ grow like $t$, and $\varepsilon_0$ would not be integrable. In the variants we tested, changing the number of terms and the fit window of the tails changes the distributions negligibly.

\begin{figure}[htb]
    \centering
    \includegraphics[width=\linewidth]{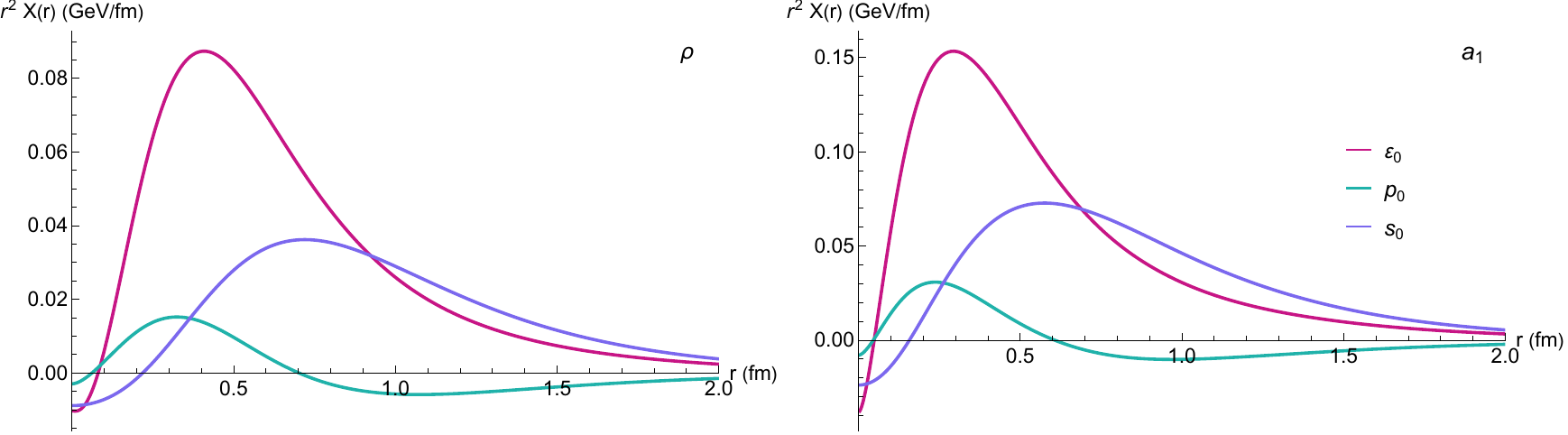}
    \caption{The $r^2$-weighted monopole distributions of energy $\varepsilon_0$, pressure $p_0$ and shear force $s_0$ for $\rho$ (left) and $a_1$ (right). The legend on the right applies to both plots.}
    \label{fig:monopoledensities}
\end{figure}

Figure~\ref{fig:monopoledensities} shows the monopole distributions. The energy distribution $r^2 \varepsilon_0$ peaks at $\approx 0.4$~fm for $\rho$ and at $\approx 0.3$~fm for $a_1$. The pressure $p_0$ is positive at small $r$ and turns negative at $\approx 0.7$~fm and $\approx 0.6$~fm, respectively, while the shear force $s_0$ peaks at $\approx 0.7$~fm and $\approx 0.6$~fm. Close to the origin, $\varepsilon_0$, $p_0$ and $s_0$ are all slightly negative. These negative regions should not be read as regions of negative energy density, pressure or shear force: Interpreting the Breit-frame Fourier transforms as static spatial densities requires that the position of the hadron can be fixed to a precision better than its size (see the discussions in~\cite{Epelbaum:2021ahi, Panteleeva:2023evj, Freese:2022yur}). The regions lie within the (reduced) Compton wavelengths $1/m_\rho \approx 0.25$~fm and $1/m_{a_1} \approx 0.16$~fm, and the sign of the distributions there therefore carries no direct physical meaning.

\begin{figure}[htb]
    \centering
    \includegraphics[width=\linewidth]{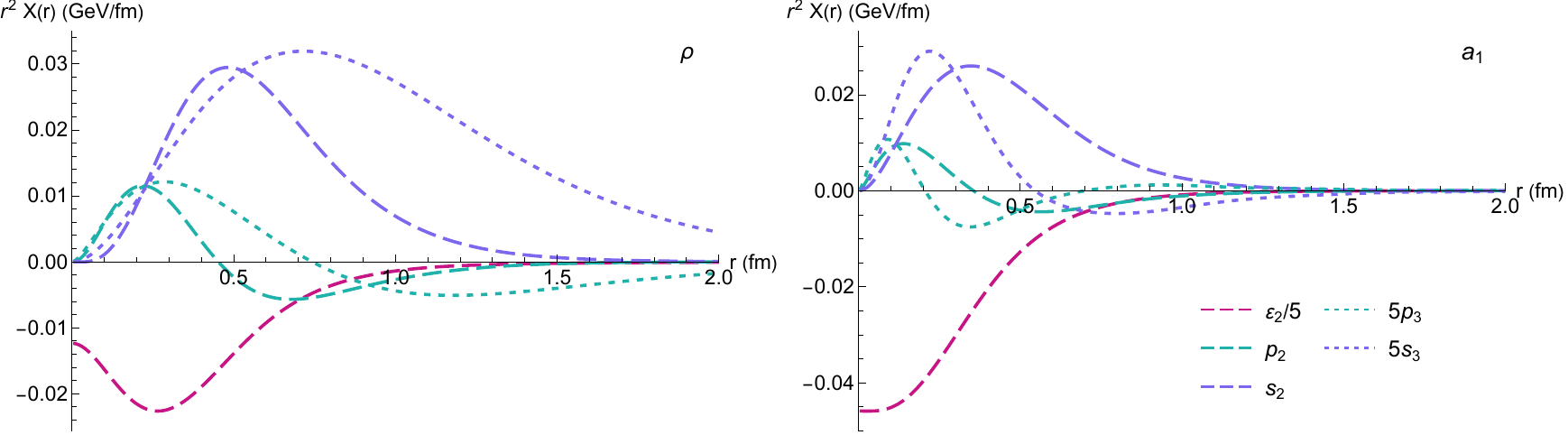}
    \caption{The $r^2$-weighted quadrupole distributions for $\rho$ (left) and $a_1$ (right): energy $\varepsilon_2$, pressure $p_2$ and shear force $s_2$, and pressure $p_3$ and shear force $s_3$, scaled as indicated for legibility. The colors denote energy, pressure and shear force as in Figure~\ref{fig:monopoledensities}, and the legend of the right panel applies to both plots.}
    \label{fig:quadrupoledensities}
\end{figure}

Figure~\ref{fig:quadrupoledensities} shows the quadrupole distributions, which would vanish for a spherically symmetric state. We have scaled $\varepsilon_2$ by $1/5$ and $p_3$ and $s_3$ by $5$ for legibility. For $a_1$, whose $\D_3$ is close to zero at small $-t$, $p_3$ and $s_3$ are the distributions most sensitive to the tails. In particular, $p_3$ and $s_3$ are peaked much closer to the center of the meson for $a_1$ than for $\rho$.

Finally, the mass and mechanical radii are defined as the second moments of the energy distribution and of the normal pressure $p_0 + \frac{2}{3}s_0$,
\begin{equation}
    \label{eq:radii}
    \expval{r^2}_\text{mass} = \frac{\int \dd^3 r r^2 \varepsilon_0}{\int \dd^3 r \varepsilon_0} \ , \qquad \expval{r^2}_\text{mech} = \frac{\int \dd^3 r r^2 \qty(p_0 + \frac{2}{3}s_0)}{\int \dd^3 r \qty(p_0 + \frac{2}{3}s_0)} \ ,
\end{equation}
where we use the monopole distributions only, since the quadrupole ones depend on the polarization. With $r_\text{mass} = \sqrt{\expval{r^2}_\text{mass}}$ and $r_\text{mech} = \sqrt{\expval{r^2}_\text{mech}}$, we find
\begin{equation}
    r_\text{mass}^\rho = 0.807 \ \text{fm} \ , \quad r_\text{mech}^\rho = 0.885 \ \text{fm} \ , \quad r_\text{mass}^{a_1} = 0.734 \ \text{fm} \ , \quad r_\text{mech}^{a_1} = 0.797 \ \text{fm} \ .
\end{equation}
The $a_1$ is more compact than $\rho$ by both measures, and for both mesons the mechanical radius exceeds the mass radius only by about $10\%$.

\section{Conclusions and outlook}
\label{sec:conclusions}

We have analyzed the structure of the $\rho$ and $a_1$ mesons in holographic V-QCD by computing the six gravitational form factors and the related mechanical properties of these mesons. Both mesons were described simultaneously by a unified spin-1 field. We extracted the matrix element of the energy-momentum tensor between spin-1 states by applying LSZ reduction to the holographic three-point function, with the normalization of the external states determined from the pole residue of the two-point function of the currents. The form factors followed from the cubic couplings of the spin-1 field to the tensor fluctuation of the metric and to the coupled scalar fluctuations of the metric, the dilaton and the tachyon.

In the tensor channel we found that $A_0 = J$ and $A_1 = 0$ hold identically in the model, and that $A_0(0) = J(0) = 1$, as required by Poincaré invariance, followed automatically from the normalization of the external states. The scalar channel gave $D_0$ and $D_1$. The expressions for $E$, $D_0$ and $D_1$ contained explicit poles at $t = 0$, but the poles canceled exactly, and all three form factors were regular. In the range we compared, $A_0$ and $J$ of $\rho$ in V-QCD decreased faster with increasing $-t$ than in the NJL model~\cite{Freese:2019bhb}, the light-cone constituent quark model~\cite{Sun:2020wfo} and the hard-wall model~\cite{Abidin:2008ku}, and $D_0$ and $D_1$ were larger in magnitude near $t = 0$ than in the NJL model and approached zero faster.

From the form factors we computed the monopole and quadrupole distributions of energy, pressure and shear force in the Breit frame. The D-term, defined as the forward value of the combination $\D_0$, was negative for both mesons, $\D_0(0) = -2.98$ for $\rho$ and $-7.43$ for $a_1$. For both mesons the pressure $p_0$ was positive at small $r$, apart from a small negative region close to the origin, and changed sign at $r \approx 0.7$~fm for $\rho$ and $r \approx 0.6$~fm for $a_1$, becoming negative at larger $r$. We obtained mass and mechanical radii of $0.807$~fm and $0.885$~fm for $\rho$, and $0.734$~fm and $0.797$~fm for $a_1$. 

Several directions lie open for further study. First, the relations $A_0 = J$ and $A_1 = 0$ most likely reflect the simplicity of the flavor action at quadratic order rather than a property of spin-1 hadrons, since the NJL and light-cone results do not satisfy them~\cite{Freese:2019bhb, Sun:2020wfo}. As for the form factor of the proton that vanished in~\cite{Hippelainen:2026izm}, further terms in the five-dimensional action could generate a non-zero $A_1$.

Second, a decomposition of the form factors into quark and gluon contributions would allow a comparison with the gluonic lattice results of~\cite{Pefkou:2021fni}. As for the pion~\cite{Hippelainen:2026pion}, such a decomposition is not straightforward in V-QCD, since the glue and flavor sectors backreact on each other.

Third, the gravitational form factors are related to moments of generalized parton distributions (GPDs)~\cite{Ji:1996ek}, also for spin-1 hadrons~\cite{Polyakov:2019lbq}, so the holographic form factors could be used to constrain parametrizations of the GPDs. A direct holographic computation of the GPDs in V-QCD could in turn test the form factors. We leave these questions for future work.

\section*{Acknowledgements}

The author thanks Roman Zwicky, Niko Jokela and Matti J\"arvinen for helpful comments and discussions during the preparation of this work.

\appendix

\section{Boundary values of the gravity fluctuations}
\label{app:boundaryvalues}

Here we relate the boundary values of the gravity fluctuations of Section~\ref{subsec:gravityfluctuations} to the source $\delta\eta_\mn$ of the energy-momentum tensor. The boundary condition for the metric is $e^{-2A}g_\mn \to \eta_\mn + \delta\eta_\mn$ as $z \to 0$, which for the parametrization~\eqref{eq:fluctuatedmetric} means
\begin{equation}
    \label{eq:boundarymetric}
    2 \psi^{(0)} \eta_\mn + 2 \pd_\mu \pd_\nu E^{(0)} + \pd_{\{\mu} V^{T(0)}_{\nu\}} + 2 h_\mn^{TT(0)} = \delta \eta_\mn \ ,
\end{equation}
where the superscript $(0)$ denotes the value of a field at $z = 0$. Contracting the Fourier transform of~\eqref{eq:boundarymetric}, with the convention~\eqref{eq:fourierconvention}, with the transverse and transverse-traceless projectors $\Theta^\mn(q)$ and $\TTproj(q)$ defined in~\eqref{eq:TTtracesplit} gives
\begin{equation}
    \label{eq:boundaryvalues}
    \widetilde{\psi}^{(0)}(q) = \frac{1}{6}\Theta^\mn(q) \widetilde{\delta \eta}_\mn(q) \ , \qquad \widetilde{h}_\mn^{TT(0)}(q) = \frac{1}{2}\TTproj(q) \widetilde{\delta \eta}_\ab(q) \ ,
\end{equation}
since $\Theta^\mn$ annihilates the terms with $\widetilde{E}^{(0)}$, $\widetilde{V}^{T(0)}_\mu$ and $\widetilde{h}^{TT(0)}_\mn$, and $\varepsilon^{TT}$ those with $\widetilde{\psi}^{(0)}$, $\widetilde{E}^{(0)}$ and $\widetilde{V}^{T(0)}_\mu$. The boundary values $E^{(0)}$ and $V^{T(0)}_\mu$ are gauge degrees of freedom and do not enter the vertices (Appendix~\ref{subapp:gaugeinvariance}).

The sources of the dilaton and of the tachyon modulus are set to zero, since they would source $\Tr G_\mn G^\mn$ and shift the quark mass. This does not make $\chi$ and $s$ vanish near the boundary, since both retain their normalizable parts, and we must check that these do not contribute to the boundary values of the invariants. The dilaton fluctuation is dual to an operator of dimension four and the tachyon fluctuation to an operator of dimension three, so near the boundary
\begin{equation}
    \label{eq:vevfalloffs}
    \chi \sim z^4 \ , \qquad s \sim z^3 \ ,
\end{equation}
up to logarithmic corrections. The background behaves as $A' \sim 1/z$, $\Phi' \sim 1/z$ and $\tau' \sim m_q$, again up to logarithms~\cite{Jarvinen:2011qe, Arean:2013tja}, where $m_q$ is the quark mass. The terms that distinguish $\zeta$ and $\varsigma$ from $\psi$ in~\eqref{eq:gaugeinvariants} therefore vanish at the boundary,
\begin{equation}
    \label{eq:vevpowercounting}
    \frac{A'}{\Phi'} \chi \sim z^4 \to 0 \ , \qquad \frac{A'}{\tau'} s \sim \frac{z^2}{m_q} \to 0 \ ,
\end{equation}
and
\begin{equation}
    \label{eq:invariantboundaryvalues}
    \zeta^{(0)} = \varsigma^{(0)} = \psi^{(0)} \ .
\end{equation}
This argument requires $m_q \neq 0$, which holds for the background we use. 

Consequently, the Fourier transform of each scalar invariant is its bulk-to-boundary propagator~\eqref{eq:propagatorboundaryvalues} multiplied by the same boundary factor, and similarly for the tensor. The fluctuations sourced by $\delta\eta_\mn$ are therefore
\begin{equation}
    \label{eq:sourcedfluctuations}
    \begin{split}
        X(x, z) &= \int \frac{\dd^4 q}{(2\pi)^4} e^{iq\cdot x} \widehat{X}(t, z) \widetilde{\psi}^{(0)}(q) \ , \qquad X \in \qty{\zeta, \varsigma, \F} \ , \\
        h^{TT}_\mn(x, z) &= \int \frac{\dd^4 q}{(2\pi)^4} e^{iq\cdot x} \widehat{h}(t, z) \widetilde{h}_\mn^{TT(0)}(q) \ ,
    \end{split}
\end{equation}
where $\widehat{\F}$ is found as the IR-regular solution of the conjugate relation~\eqref{eq:conjugateequation}, defined by $\widehat{\zeta}$ and $\widehat{\varsigma}$.

\section{Quadratic action and cubic vertices}
\label{subapp:quadraticactionandcubicvertices}

In this Appendix we derive the quadratic action of Section~\ref{subsec:vectorandaxialvectormesons} and the cubic vertices of Sections~\ref{subsec:tensorvertex} and~\ref{subsec:scalarvertex}. The derivations follow those of the pion vertices in~\cite{Hippelainen:2026pion}.

Integrations by parts in $z$ produce boundary terms, which we evaluate with the following asymptotics. In the UV, $b_1$, $b_2$, $c_1$ and $c_2$ of~\eqref{eq:b1b2c1} are $\oford{1/z}$ and the normalizable mode is $\varphi = \oford{z^2}$, all up to logarithmic corrections~\cite{Arean:2013tja}. In the IR, $V_f$ contains the factor $e^{-\widetilde{a}_2 \tau^2}$ with a constant $\widetilde{a}_2 > 0$ and $\tau$ growing as a power of $z$~\cite{Amorim:2021gat}, so $b_1$, $b_2$, $c_1$ and $c_2$ vanish faster than any power of $z$, while the IR-regular modes grow at most as a power.

\subsection{Quadratic action}
\label{subapp:quadraticaction}

With $T = \alphasymbol e^{i\Pi}$ and to linear order in $\Pi$, $L_M$ and $R_M$, the covariant derivative~\eqref{eq:covariantderivative} is $D_M T = \pd_M \alphasymbol + i \alphasymbol \qty(\pd_M \Pi + 2A_M)$, so the matrices~\eqref{eq:dbimatrices} become
\begin{equation}
    \label{eq:dbimatricesexpanded}
    \boldsymbol{L}_\MN = \widetilde{g}_\MN + w F^L_\MN + \kappa \alphasymbol^2 \qty(\pd_M \Pi + 2A_M)\qty(\pd_N \Pi + 2A_N) \ ,
\end{equation}
and $\boldsymbol{R}_\MN$ is the same with $F^L_\MN \to F^R_\MN$. Expanding the determinants to second order in the fluctuations around $\widetilde{g}_\MN$, the term linear in the antisymmetric field strength vanishes, and using $(F^L)^2 + (F^R)^2 = 2 (F^V)^2 + 2 (F^A)^2$ for the Abelian parts gives~\eqref{eq:quadraticflavoraction}.

On the background, with $A_\mu = A^\perp_\mu + A^\parallel_\mu$, the field-strength and tachyon terms of the axial field are
\begin{equation}
    \label{eq:axialopened}
    \begin{split}
        F^A_\MN \widetilde{g}^{MO}\widetilde{g}^{NP}F^A_{OP} &= e^{-4A}\qty[A^\perp_\mn A^{\perp\mn} + \frac{2}{G^2}\qty(\pd_z A^\perp_\mu - \pd_\mu A_z + \pd_z A^\parallel_\mu)^2] \ , \\
        \widetilde{g}^\MN \qty(\pd_M \Pi + 2A_M)\qty(\pd_N \Pi + 2A_N) &= e^{-2A}\qty(2 A^\perp_\mu + \pd_\mu \Pi + 2 A^\parallel_\mu)^2 + \frac{e^{-2A}}{G^2}\qty(\Pi' + 2A_z)^2 \ ,
    \end{split}
\end{equation}
with squares taken with $\eta^\mn$. After opening the squares and integrating by parts, every cross term between $A^\perp_\mu$ and $(\Pi, A^\parallel_\mu, A_z)$ vanishes. The transverse axial field therefore decouples, and inserting the remaining terms into~\eqref{eq:quadraticflavoraction} gives the axial action of~\eqref{eq:quadraticactions} with the factors~\eqref{eq:b1b2c1}. The vector action follows analogously, without the tachyon term.

For the unified field $B_\mu$ in the gauge $B_z = 0$, the action on the background can be written as
\begin{equation}
    \label{eq:L0T}
    S^{(2)} = -\frac{1}{2}M^3 N_c \Tr \int \dd^5 x \L^{(0)} \ , \qquad \L^{(0)} = \frac{b_1}{2}\sfY + b_2 \sfX + 4c_1 \sfM \ ,
\end{equation}
with the factors $\sfX_\mn$, $\sfY_\mn$ and $\sfM_\mn$ of~\eqref{eq:tensorvertex} and their traces. We also denote
\begin{equation}
    \label{eq:TmnEmu}
    \Delta_\mu = b_1 \Box B_\mu + \qty(b_2 B_\mu')' - 4c_1 B_\mu \ ,
\end{equation}
where $\Delta_\mu = 0$ is the equation of motion~\eqref{eq:eomsgeneral} of the unified field. In the following, we keep $\Delta_\mu$ explicit, so that the places where the equation of motion is used are visible.

\subsection{Tensor vertex}
\label{subapp:tensorvertexderivation}

The cubic vertices are the first-order fluctuations of~\eqref{eq:quadraticflavoraction}, with $\pd_M \Pi + 2A_M \to 2B_M$, with respect to the metric, the dilaton and the tachyon. For the tensor fluctuation, $\delta\widetilde{g}^\mn = -2e^{-2A}h^{TT,\mn}$, the measure does not change, and with $F_\MN$ the field strength of $B_M$,
\begin{equation}
    \label{eq:tensorfluctuations}
    \begin{split}
        \delta\qty(F_\MN \widetilde{g}^{MO}\widetilde{g}^{NP}F_{OP}) &= -4e^{-4A} h^{TT,\mn}\qty(\sfY_\mn + \frac{1}{G^2}\sfX_\mn) \ , \\
        \delta\qty(4 \widetilde{g}^\MN B_M B_N) &= -8 e^{-2A} h^{TT,\mn}\sfM_\mn \ .
    \end{split}
\end{equation}
Summing these and multiplying by the prefactors of~\eqref{eq:quadraticflavoraction} gives~\eqref{eq:tensorvertex}.

\subsection{Scalar vertex}
\label{subapp:scalarvertexderivation}

The scalar and transverse vector parts of~\eqref{eq:fluctuatedmetric}, together with $\chi$ and $s$, fluctuate the effective metric by
\begin{equation}
    \label{eq:effectivemetricfluctuationsVA}
    \begin{split}
        \delta\widetilde{g}_\mn &= e^{2A}\qty(2\psi \eta_\mn + 2\pd_\mu \pd_\nu E + \pd_{\{\mu}V^T_{\nu\}}) \ , \\
        \delta\widetilde{g}_{\mu z} &= e^{2A}\qty(\pd_\mu W + A^T_\mu + e^{-2A}\kappa \tau' \pd_\mu s) \ , \\
        \delta\widetilde{g}_{zz} &= 2e^{2A} \Xi \ , \qquad \Xi = \phi + e^{-2A}\kappa \tau' s' + \frac{1}{2}e^{-2A} \qty(\tau')^2 \chi \pd_\Phi \kappa \ .
    \end{split}
\end{equation}
The measure, the inverse effective metric and the potentials fluctuate as
\begin{equation}
    \label{eq:measureandinversefluctuations}
    \begin{split}
    \delta \sqrt{-\det \widetilde{g}} &= e^{5A} G \qty(4 \psi + \Box E + \pd \cdot V^T + \frac{\Xi}{G^2}) \ , \\
    \delta \widetilde{g}^\mn &= - e^{-2A} \qty(2\psi \eta^\mn + 2\pd^\mu \pd^\nu E + \pd^{\{\mu}V^{T,\nu\}}) \ , \\
    \delta \widetilde{g}^{\mu z} &= -\frac{e^{-2A}}{G^2} \qty(\pd^\mu W + A^{T,\mu} + e^{-2A} \kappa \tau' \pd^\mu s) \ , \qquad
    \delta \widetilde{g}^{zz} = -\frac{2 e^{-2A}}{G^4} \Xi \ , \\
    \delta V_f &= \chi \pd_\Phi V_f + s \pd_\tau V_f \ , \qquad
    \delta \qty(\kappa \alphasymbol^2) = \tau^2 \chi \pd_\Phi \kappa + 2 \kappa \tau s \ , \qquad
    \delta \qty(w^2) = 2 w \chi \pd_\Phi w \ ,
    \end{split}
\end{equation}
with all potentials and their derivatives evaluated on the background. Even though $\pd \cdot V^T = 0$, we have kept it in the fluctuation of the determinant, since this allows us to treat $E$ and $V_\mu^T$ together. Collecting all terms, the first-order fluctuation before any choice of gauge is
\begin{equation}
    \label{eq:masterfluctuationVA}
    \begin{split}
        S^{(3)}_{sBB} = -\frac{1}{2}M^3 N_c \Tr \int \dd^5 x \Big\{
        &\Xi\qty[\frac{b_2}{2}\sfY - \frac{b_2}{G^2}\sfX + \frac{4c_1}{G^2}\sfM] + \psi\qty(2b_2 \sfX + 8c_1 \sfM) \\
        &+ \chi\qty[\qty(\frac{\pd_\Phi V_f}{V_f} + \frac{2\pd_\Phi w}{w})\qty(\frac{b_1}{2}\sfY + b_2\sfX) + 4c_1\qty(\frac{\pd_\Phi V_f}{V_f} + \frac{\pd_\Phi \kappa}{\kappa})\sfM] \\
        &+ s\qty[\frac{\pd_\tau V_f}{V_f}\qty(\frac{b_1}{2}\sfY + b_2\sfX) + 4c_1\qty(\frac{\pd_\tau V_f}{V_f} + \frac{2}{\tau})\sfM] \\
        &+ \qty(\pd^\mu \vartheta_\mu)\L^{(0)} - 2\qty(\pd^\mu \vartheta^\nu)\qty(b_1 \sfY_\mn + b_2 \sfX_\mn + 4c_1 \sfM_\mn) \\
        &- 2b_2\qty(\pd^\mu W + A^{T,\mu} + e^{-2A}\kappa \tau' \pd^\mu s)\sfZ_\mu \Big\} \ ,
    \end{split}
\end{equation}
where for the purposes of the gauge invariance check, we already wrote $\vartheta_\mu = \pd_\mu E + V^T_\mu$. We show in Appendix~\ref{subapp:gaugeinvariance} that the gauge modes $\chi$, $E$ and $V^T_\mu$ drop out of~\eqref{eq:masterfluctuationVA} on-shell, and therefore set $\chi = E = V^T_\mu = 0$ in the rest of this subsection. In this gauge, the inverted relations~\eqref{eq:invertedscalars} give
\begin{equation}
    \label{eq:gaugefixedVA}
    \begin{split}
        \psi &= \zeta \ , \quad A^T_\mu = \W^T_\mu \ , \quad s = \frac{\tau'}{A'}\qty(\zeta - \varsigma) \ , \quad W = \frac{\zeta}{A'} - \F \ , \\
        \frac{\Xi}{G^2} &= \sigma\qty(\zeta - \varsigma) + \frac{\zeta'}{A'} - \frac{G^2 - 1}{G^2 A'}\varsigma' \ , \quad
        W + e^{-2A}\kappa \tau' s = \frac{G^2 \zeta - (G^2 - 1)\varsigma}{A'}- \F \ ,
    \end{split}
\end{equation}
with $\sigma$ from~\eqref{eq:sigmavtau}, and where we used $e^{-2A}\kappa (\tau')^2 = G^2 - 1$. To bring the vertex to its canonical form, we first integrate $\pd^\mu$ by parts off the scalars in the last term of~\eqref{eq:masterfluctuationVA} and reduce the result with
\begin{equation}
    \label{eq:divZ}
    \pd^\mu \sfZ_\mu = \frac{1}{4}\pd_z \sfY - \frac{1}{b_1}\qty[b_2' \sfX + \frac{b_2}{2}\pd_z \sfX - 2c_1 \pd_z \sfM] + \frac{1}{b_1}\Delta^\nu B'_\nu \ ,
\end{equation}
which follows from $\pd^\mu B_\mn = \Box B_\nu$ for a transverse field, the antisymmetry of $B_\mn$, and the definition of $\Delta_\mu$ in~\eqref{eq:TmnEmu}. On-shell, the vertex then reads
\begin{equation}
\label{eq:scalarvertexintermediate}
\begin{split}
    S^{(3)}_{sBB} &= -\frac{1}{2}M^3 N_c \Tr \int \dd^5 x \Bigg\{ -2 b_2 \W^{T, \mu}\sfZ_\mu \\
    &+\qty(\sigma(\zeta-\varsigma) + \frac{\zeta'}{A'} - \frac{G^2 -1}{G^2 A'}\varsigma')\qty(\frac{b_1}{2}\sfY - b_2 \sfX + 4 c_1 \sfM) \\
    &+\zeta\qty(2 b_2 \sfX + 8 c_1 \sfM) + \frac{\tau'}{A'}(\zeta - \varsigma)\qty[\frac{\pd_\tau V_f}{V_f} \qty(\frac{b_1}{2}\sfY + b_2 \sfX) + 4 c_1 \qty(\frac{\pd_\tau V_f}{V_f} + \frac{2}{\tau})\sfM] \\
    &+2b_2\qty(\frac{G^2 \zeta - (G^2-1)\varsigma}{A'} - \F) \qty(\frac{1}{4}\pd_z \sfY - \frac{1}{b_1}\qty(b_2' \sfX + \frac{b_2}{2}\pd_z \sfX) + \frac{2c_1}{b_1}\pd_z \sfM) \Bigg\} \ .
\end{split}
\end{equation}
Finally, we integrate the radial derivatives by parts off $\sfX, \sfY$ and $\sfM$ in the last line, and eliminate $\F'$ with the conjugate relation~\eqref{eq:conjugateequation}. The terms with $\zeta'$ and $\varsigma'$ cancel, and collecting the coefficients of the nine products of $\qty{\sfY, \sfX, \sfM}$ with $\qty{\zeta, \varsigma, \F}$ gives~\eqref{eq:scalarvertex} with the coefficients~\eqref{eq:scalarcoefficients}. The boundary terms of the radial integrations by parts are proportional to $b_2 \sfY$, $b_2 \sfX$ or $c_1 \sfM$ evaluated on the normalizable modes, and vanish both in the UV and in the IR by the asymptotics given at the beginning of this Appendix.

\subsection{Gauge invariance of the scalar vertex}
\label{subapp:gaugeinvariance}

Here we show that $\chi$, $E$ and $V^T_\mu$ drop out of~\eqref{eq:masterfluctuationVA} once the equation of motion of $B_\mu$ is used. We need the following identities, which follow from the antisymmetry of $B_\mn$ and from $\pd^\mu B_{\mn} = \Box B_\nu$ for a transverse field,
\begin{equation}
    \label{eq:bilinearidentities}
    \pd^\mu \sfY_\mn = \qty(\Box B^\rho) B_{\nu\rho} + \frac{1}{4}\pd_\nu \sfY \ , \quad \pd^\mu \sfX_\mn = \frac{1}{2}\pd_\nu \sfX + B'^\mu \pd_z B_\mn \ , \quad \pd^\mu \sfM_\mn = \frac{1}{2}\pd_\nu \sfM + B^\mu B_\mn \ .
\end{equation}
Together with the definitions~\eqref{eq:L0T} and~\eqref{eq:TmnEmu} they imply
\begin{equation}
    \label{eq:Tidentity}
    2\pd^\mu \qty(b_1 \sfY_\mn + b_2 \sfX_\mn + 4c_1 \sfM_\mn) + 2\pd_z\qty(b_2 \sfZ_\nu) = \pd_\nu \L^{(0)} + 2\Delta^\rho B_{\nu\rho} \ .
\end{equation}

The fields $E$ and $V_\mu^T$ enter~\eqref{eq:masterfluctuationVA} both explicitly and through $W$ and $A^{T}_\mu$ in the last line, by~\eqref{eq:invertedscalars}, but only in the combination $\vartheta_\mu = \pd_\mu E + V^T_\mu$. The terms containing $\vartheta_\mu$ are
\begin{equation}
    \label{eq:varthetaterms}
    \qty(\pd^\mu \vartheta_\mu)\L^{(0)} - 2\qty(\pd^\mu \vartheta^\nu)\qty(b_1 \sfY_\mn + b_2 \sfX_\mn + 4c_1 \sfM_\mn) - 2b_2 (\vartheta^\mu)' \sfZ_\mu \ .
\end{equation}
Integrating by parts and using~\eqref{eq:Tidentity}, we find
\begin{equation}
    \label{eq:varthetatermsreduced}
    \vartheta^\nu\qty[-\pd_\nu \L^{(0)} + 2\pd^\mu \qty(b_1 \sfY_\mn + b_2 \sfX_\mn + 4c_1 \sfM_\mn) + 2\pd_z\qty(b_2 \sfZ_\nu)] = 2\vartheta^\nu \Delta^\rho B_{\nu\rho} \ ,
\end{equation}
which vanishes on-shell. The radial boundary term $-2b_2\vartheta^\mu \sfZ_\mu$ vanishes in the UV, where $b_2 \sfZ_\mu = \oford{z^2}$ on the normalizable modes, and in the IR, where $b_2$ vanishes.

For the $\chi$ gauge invariance check, we need to extract all $\chi$-dependent parts of the fluctuations in~\eqref{eq:masterfluctuationVA}. Only for this computation, by~\eqref{eq:invertedscalars}, we may replace
\begin{equation}
    \label{eq:chiparts}
    \begin{split}
        \psi &\to A'\overline{\chi} \ , \qquad s \to \tau'\overline{\chi} \ , \qquad W \to \overline{\chi} \ , \qquad \phi \to \overline{\chi}' + A'\overline{\chi} \ , \\
        \frac{\Xi}{G^2} &\to \overline{\chi}' + \qty(A' + \frac{G'}{G})\overline{\chi} \ , \qquad W + e^{-2A}\kappa \tau' s \to G^2 \overline{\chi} \ ,
    \end{split}
\end{equation}
where for simplicity we defined $\overline{\chi} = \chi/\Phi'$. We denote the radial derivatives of the potentials on the background by $V_f' = \Phi' \pd_\Phi V_f + \tau' \pd_\tau V_f$, $w' = \Phi' \pd_\Phi w$ and $\kappa' = \Phi' \pd_\Phi \kappa$. After the last term of~\eqref{eq:masterfluctuationVA} is integrated by parts and reduced on-shell with~\eqref{eq:divZ}, the complete $\overline{\chi}$-dependence of the integrand is
\begin{equation}
    \label{eq:varepsilondependence}
    \begin{split}
        &\qty[\overline{\chi}' + \qty(A' + \frac{G'}{G})\overline{\chi}]\qty(\frac{b_1}{2}\sfY - b_2 \sfX + 4c_1\sfM) + A' \overline{\chi}\qty(2 b_2 \sfX + 8 c_1 \sfM) \\
        &+ \overline{\chi} \qty[\qty(\frac{V_f'}{V_f} + \frac{2 w'}{w})\qty(\frac{b_1}{2} \sfY + b_2 \sfX) + 4 c_1 \qty(\frac{V_f'}{V_f} + \frac{\kappa'}{\kappa} + \frac{2\tau'}{\tau})\sfM] \\
        &+2 b_1 \overline{\chi} \qty[\frac{1}{4}\pd_z \sfY - \frac{1}{b_1}\qty(b_2' \sfX + \frac{b_2}{2}\pd_z \sfX) + \frac{2 c_1}{b_1}\pd_z \sfM] \ .
    \end{split}
\end{equation}
Integrating the radial derivatives by parts off the traced bilinears, the last line becomes
\begin{equation}
    \label{eq:chilastline}
    -\frac{1}{2}\qty(b_1 \overline{\chi})' \sfY - b_2' \overline{\chi} \sfX + b_2 \overline{\chi}' \sfX - 4 \qty(c_1 \overline{\chi})' \sfM \ ,
\end{equation}
and the terms proportional to $\overline{\chi}'$ cancel against those of the first line of~\eqref{eq:varepsilondependence}. The remaining terms are $\overline{\chi}$ times
\begin{equation}
    \label{eq:chicoefficients}
    \begin{split}
        &\frac{b_1}{2}\qty[A' + \frac{G'}{G} + \frac{V_f'}{V_f} + \frac{2 w'}{w} - \frac{b_1'}{b_1}]\sfY + b_2 \qty[A' - \frac{G'}{G} + \frac{V_f'}{V_f} + \frac{2 w'}{w} - \frac{b_2'}{b_2}]\sfX \\
        &+ 4c_1 \qty[3A' + \frac{G'}{G} + \frac{V_f'}{V_f} + \frac{\kappa'}{\kappa} + \frac{2\tau'}{\tau} - \frac{c_1'}{c_1}]\sfM \ ,
    \end{split}
\end{equation}
and each bracket vanishes by the definitions~\eqref{eq:b1b2c1}. The boundary terms again vanish in the UV and in the IR, and the full vertex is gauge invariant.

\section{Regularity of the form factors at \texorpdfstring{$t = 0$}{t = 0}}
\label{subapp:regularity}

The form factors~\eqref{eq:tensorFFs} and~\eqref{eq:D0D1} contain explicit poles at $t = 0$. Expanding the integrals around $t = 0$, all of them cancel if
\begin{equation}
    \label{eq:regularityconditions}
    \begin{split}
    m^2 \T_\sfY(0) - \T_\sfX(0) - \T_\sfM(0) &= 0 \ , \\
     \S_\sfX(0) + \S_\sfM(0) - m^2 \S_\sfY(0) &= 4m^2 \ , \\
      -8 \pd_{t} \qty(m^2 \T_\sfY - \T_\sfX - \T_\sfM) \eval_{t = 0} &= \S_\sfY(0) \ .
    \end{split}
\end{equation}
The first condition removes the pole of $E$ and the $1/t^2$ pole of $D_1$, the second the pole of $D_0$, and the third the remaining $1/t$ pole of $D_1$. In this Appendix we show that all three hold. The argument for the second and the third condition parallels that for the pion~\cite{Hippelainen:2026pion}.

The first condition only requires the tensor propagator at $t = 0$. There the general solution of~\eqref{eq:tensorpropagator} is $\widehat{h} = C_1 + C_2 \int_0^z \dd u e^{-3A(u)}$ with constants $C_1$ and $C_2$, and regularity in the IR together with the boundary condition leaves
\begin{equation}
    \label{eq:h0}
    \widehat{h}(0, z) = 1 \ .
\end{equation}
The tensor integrals~\eqref{eq:tensorintegrals} then reduce to integrals of the profile alone: $\T_\sfY(0) = 1$ by the definition of $\N$ in~\eqref{eq:canonicalaction}, and $\T_\sfX(0) + \T_\sfM(0) = m^2$ by~\eqref{eq:massidentity}. This is the first condition, so the bracket multiplying $1/t$ in $E$ of~\eqref{eq:tensorFFs} vanishes at $t = 0$ and $E$ is finite. By~\eqref{eq:tensorFFs}, the same relations give $A_0(0) = J(0) = 1$ for any normalization of $\varphi$. The values fixed by Poincar\'e symmetry are thus reproduced, which is a cross-check of the normalization of the external states.

For the other two conditions we need the propagators at small $t$. At $t = 0$ the coupled scalar equations are solved by $\widehat{\zeta} = \widehat{\varsigma} = 1$~\cite{Hippelainen:2026pion}, and by~\eqref{eq:conjugateequation} and~\eqref{eq:eeconstraint} the function $\F_0(z) = \widehat{\F}(0, z)$ satisfies
\begin{equation}
    \label{eq:F0equation}
    \F_0' + 3A' \F_0 = 1 - \frac{A''}{(A')^2} \ .
\end{equation}
The shifted function $Y = \F_0 - 1/A'$ then satisfies
\begin{equation}
    \label{eq:Yequation}
    Y' + 3A'Y = -2 \ , \qquad Y(z) = 2e^{-3A(z)}\int_z^\infty \dd u e^{3A(u)} \ ,
\end{equation}
where the homogeneous solution $e^{-3A}$ is excluded by regularity in the IR. Up to logarithmic corrections, $Y \sim z$ in the UV, and $Y \simeq -2/(3A') \sim 1/z$ in the IR. 

For the third condition, we expand the tensor propagator as $\widehat{h}(t, z) = 1 + t h_1(z) + \oford{t^2}$ with the leading term~\eqref{eq:h0}, and~\eqref{eq:tensorpropagator} gives $h_1'' + 3A' h_1' + 1 = 0$ with $h_1(0) = 0$. The IR-regular solution has
\begin{equation}
    \label{eq:h1}
    h_1' = e^{-3A(z)}\int_z^\infty \dd u e^{3A(u)} = \frac{Y}{2} \ .
\end{equation}

Next we evaluate the combinations $\C_I(0, z)$ of~\eqref{eq:scriptC}. With $\widehat{\zeta} = \widehat{\varsigma} = 1$ they contain the sums of the coefficients of $\zeta$ and $\varsigma$ in~\eqref{eq:scalarcoefficients}, which simplify with~\eqref{eq:eeconstraint} to
\begin{equation}
    \label{eq:coefficientsums}
    \begin{split}
        c_{\sfY\zeta} + c_{\sfY\varsigma} &= \frac{b_2}{2}\qty(1 - \frac{A''}{(A')^2}) - \qty(\frac{b_2}{2A'})' \ , \\
        c_{\sfX\zeta} + c_{\sfX\varsigma} &= 2b_2 - \frac{b_2}{G^2}\qty(1 - \frac{A''}{(A')^2}) + \qty(\frac{b_2}{G^2 A'})' - \frac{2b_2'}{G^2 A'} \ , \\
        c_{\sfM\zeta} + c_{\sfM\varsigma} &= 8c_1 + 4c_2\qty(1 - \frac{A''}{(A')^2}) - \qty(\frac{4c_2}{A'})' \ .
    \end{split}
\end{equation}
The coefficients of $\F$ in~\eqref{eq:scalarcoefficients} are $c_{\sfY\F} = f' - 3A'f$ with $f = b_2/2$, $c_{\sfX\F} = f' - 3A'f + 2b_2'/G^2$ with $f = -b_2/G^2$, and $c_{\sfM\F} = f' - 3A'f$ with $f = 4c_2$. For any function $f(z)$,~\eqref{eq:F0equation} implies
\begin{equation}
    \label{eq:fF0identity}
    (f' - 3 A' f)\F_0 = \qty(f \F_0)' - f \qty(1 - \frac{A''}{(A')^2}) \ ,
\end{equation}
and using this for each of the three coefficients we find
\begin{equation}
    \label{eq:C0}
    \C_\sfY(0, z) = \frac{1}{2}\qty(b_2 Y)' \ , \quad \C_\sfX(0, z) = 2b_2 - \qty(\frac{b_2 Y}{G^2})' + \frac{2b_2'}{G^2}Y \ , \quad \C_\sfM(0, z) = 8c_1 + 4\qty(c_2 Y)' \ .
\end{equation}

For the second condition in~\eqref{eq:regularityconditions}, we insert~\eqref{eq:C0} into~\eqref{eq:scalarintegrals},
\begin{equation}
    \label{eq:secondcondition}
    \begin{split}
        &\S_\sfX(0) + \S_\sfM(0) - m^2\S_\sfY(0) = \frac{M^3 N_c}{\N}\int \dd z\qty[2b_2\qty(\varphi')^2 + 8c_1\varphi^2] \\
        &\qquad + \frac{M^3 N_c}{\N}\int \dd z\qty[\qty(\frac{2b_2'}{G^2}Y - \qty(\frac{b_2 Y}{G^2})')\qty(\varphi')^2 + \qty(4\qty(c_2 Y)' - m^2\qty(b_2 Y)')\varphi^2] \ .
    \end{split}
\end{equation}
By~\eqref{eq:massidentity}, the first integral gives $4m^2$. In the second integral we integrate by parts the terms with derivatives of $Y$, and it becomes
\begin{equation}
    \label{eq:secondconditionremainder}
    \frac{M^3 N_c}{\N}\int \dd z \frac{2 Y \varphi'}{G^2}\qty[b_2'\varphi' + b_2\varphi'' - 4c_1\varphi + m^2 b_1 \varphi] \ ,
\end{equation}
where we used $c_2 = c_1/G^2$ and $b_2 = b_1/G^2$. The bracket vanishes by the mode equation~\eqref{eq:modeequation}. The boundary terms are $\oford{z^2}$ in the UV and vanish in the IR with $b_2$ and $c_2$. This proves the second condition.

For the third condition, all $t$-dependence of the tensor integrals~\eqref{eq:tensorintegrals} is in $\widehat{h}$, so
\begin{equation}
    \label{eq:thirdconditionstep}
    \begin{split}
        \pd_{t}\qty(m^2 \T_\sfY - \T_\sfX - \T_\sfM)\eval_{t = 0} &= \frac{M^3 N_c}{2\N} \int \dd z h_1\qty[m^2 b_1 \varphi^2 - 4 c_1 \varphi^2 - b_2 (\varphi')^2] \\
        &= -\frac{M^3 N_c}{2\N} \int \dd z h_1 \qty(b_2 \varphi \varphi')' \ ,
    \end{split}
\end{equation}
where the second equality uses~\eqref{eq:modeequation}. Integrating by parts with~\eqref{eq:h1}, where the boundary terms vanish since $h_1(0) = 0$ and $b_2$ vanishes in the IR, we get
\begin{equation}
    \label{eq:thirdconditionresult}
    -8\pd_{t}\qty(m^2 \T_\sfY - \T_\sfX - \T_\sfM)\eval_{t = 0} = -\frac{2 M^3 N_c}{\N} \int \dd z Y b_2 \varphi \varphi' \ .
\end{equation}
On the other hand, inserting $\C_\sfY(0, z)$ of~\eqref{eq:C0} into~\eqref{eq:scalarintegrals} and integrating by parts once gives the same expression for $\S_\sfY(0)$, which proves the third condition. Hence $E$, $D_0$ and $D_1$ are all regular at $t = 0$. For $E$ the limit can be given in closed form: by~\eqref{eq:tensorFFs} and~\eqref{eq:thirdconditionresult},
\begin{equation}
    \label{eq:E0closedform}
    E(0) = 1 + \frac{1}{4}\S_\sfY(0) = 1 - \frac{M^3 N_c}{2\N} \int \dd z Y b_2 \varphi \varphi' \ .
\end{equation}

\clearpage

\bibliographystyle{JHEP}
\bibliography{bibliography}

@article{Cotogno:2019vjb,
    author = "Cotogno, Sabrina and Lorc{\'e}, C{\'e}dric and Lowdon, Peter and Morales, Manuel",
    title = "{Covariant multipole expansion of local currents for massive states of any spin}",
    eprint = "1912.08749",
    archivePrefix = "arXiv",
    primaryClass = "hep-ph",
    doi = "10.1103/PhysRevD.101.056016",
    journal = "Phys. Rev. D",
    volume = "101",
    number = "5",
    pages = "056016",
    year = "2020"
}

@article{Polyakov:2019lbq,
    author = "Polyakov, Maxim V. and Sun, Bao-Dong",
    title = "{Gravitational form factors of a spin one particle}",
    eprint = "1903.02738",
    archivePrefix = "arXiv",
    primaryClass = "hep-ph",
    doi = "10.1103/PhysRevD.100.036003",
    journal = "Phys. Rev. D",
    volume = "100",
    number = "3",
    pages = "036003",
    year = "2019"
}

@article{Epelbaum:2021ahi,
    author = "Epelbaum, E. and Gegelia, J. and Mei{\ss}ner, U. -G. and Polyakov, M. V.",
    title = "{Chiral theory of {\ensuremath{\rho}}-meson gravitational form factors}",
    eprint = "2109.10826",
    archivePrefix = "arXiv",
    primaryClass = "hep-ph",
    doi = "10.1103/PhysRevD.105.016018",
    journal = "Phys. Rev. D",
    volume = "105",
    number = "1",
    pages = "016018",
    year = "2022"
}

@article{Pefkou:2021fni,
    author = "Pefkou, Dimitra A. and Hackett, Daniel C. and Shanahan, Phiala E.",
    title = "{Gluon gravitational structure of hadrons of different spin}",
    eprint = "2107.10368",
    archivePrefix = "arXiv",
    primaryClass = "hep-lat",
    reportNumber = "MIT-CTP/5318",
    doi = "10.1103/PhysRevD.105.054509",
    journal = "Phys. Rev. D",
    volume = "105",
    number = "5",
    pages = "054509",
    year = "2022"
}

@article{Aliev:2020aih,
    author = "Aliev, T. M. and {\c{S}}im{\c{s}}ek, K. and Barakat, T.",
    title = "{Gravitational formfactors of the $\rho$, $\pi$, and $K$ mesons in the light-cone QCD sum rules}",
    eprint = "2009.07926",
    archivePrefix = "arXiv",
    primaryClass = "hep-ph",
    doi = "10.1103/PhysRevD.103.054001",
    journal = "Phys. Rev. D",
    volume = "103",
    pages = "054001",
    year = "2021"
}

@article{Gursoy:2007cb,
    author = "Gursoy, U. and Kiritsis, E.",
    title = "{Exploring improved holographic theories for QCD: Part I}",
    eprint = "0707.1324",
    archivePrefix = "arXiv",
    primaryClass = "hep-th",
    reportNumber = "CPHT-RR027-0507",
    doi = "10.1088/1126-6708/2008/02/032",
    journal = "JHEP",
    volume = "02",
    pages = "032",
    year = "2008"
}

@article{Gursoy:2007er,
    author = "Gursoy, U. and Kiritsis, E. and Nitti, F.",
    title = "{Exploring improved holographic theories for QCD: Part II}",
    eprint = "0707.1349",
    archivePrefix = "arXiv",
    primaryClass = "hep-th",
    reportNumber = "CPHT-RR028-0507",
    doi = "10.1088/1126-6708/2008/02/019",
    journal = "JHEP",
    volume = "02",
    pages = "019",
    year = "2008"
}

@article{Amorim:2021gat,
    author = {Amorim, Artur and Costa, Miguel S. and J{\"a}rvinen, Matti},
    title = "{Regge theory in a holographic dual of QCD in the Veneziano limit}",
    eprint = "2102.11296",
    archivePrefix = "arXiv",
    primaryClass = "hep-ph",
    reportNumber = "APCTP Pre2021 - 002",
    doi = "10.1007/JHEP07(2021)065",
    journal = "JHEP",
    volume = "07",
    pages = "065",
    year = "2021"
}

@article{Jarvinen:2011qe,
    author = "Jarvinen, Matti and Kiritsis, Elias",
    title = "{Holographic Models for QCD in the Veneziano Limit}",
    eprint = "1112.1261",
    archivePrefix = "arXiv",
    primaryClass = "hep-ph",
    reportNumber = "CCTP-2011-30",
    doi = "10.1007/JHEP03(2012)002",
    journal = "JHEP",
    volume = "03",
    pages = "002",
    year = "2012"
}

@article{Arean:2013tja,
    author = {Are{\'a}n, Daniel and Iatrakis, Ioannis and J{\"a}rvinen, Matti and Kiritsis, Elias},
    title = "{The discontinuities of conformal transitions and mass spectra of V-QCD}",
    eprint = "1309.2286",
    archivePrefix = "arXiv",
    primaryClass = "hep-ph",
    reportNumber = "CCTP-2013-7, CERN-PH-TH-2013-266",
    doi = "10.1007/JHEP11(2013)068",
    journal = "JHEP",
    volume = "11",
    pages = "068",
    year = "2013"
}

@article{Abidin:2008ku,
    author = "Abidin, Zainul and Carlson, Carl E.",
    title = "{Gravitational form factors of vector mesons in an AdS/QCD model}",
    eprint = "0801.3839",
    archivePrefix = "arXiv",
    primaryClass = "hep-ph",
    doi = "10.1103/PhysRevD.77.095007",
    journal = "Phys. Rev. D",
    volume = "77",
    pages = "095007",
    year = "2008"
}

@article{Abidin:2008hn,
    author = "Abidin, Zainul and Carlson, Carl E.",
    title = "{Gravitational Form Factors in the Axial Sector from an AdS/QCD Model}",
    eprint = "0804.0214",
    archivePrefix = "arXiv",
    primaryClass = "hep-ph",
    doi = "10.1103/PhysRevD.77.115021",
    journal = "Phys. Rev. D",
    volume = "77",
    pages = "115021",
    year = "2008"
}

@article{Hippelainen:2026izm,
    author = {Hippel{\"a}inen, Antti and Jokela, Niko and J{\"a}rvinen, Matti},
    title = "{Gravitational form factors of the proton in the improved holographic QCD model}",
    eprint = "2601.14939",
    archivePrefix = "arXiv",
    primaryClass = "hep-th",
    reportNumber = "HIP-2026-1/TH, APCTP Pre2026 - 001",
    doi = "10.1007/JHEP11(2026)061",
    journal = "JHEP",
    volume = "11",
    pages = "061",
    year = "2026"
}

@article{Kiritsis:2006ua,
    author = "Kiritsis, Elias and Nitti, F.",
    title = "{On massless 4D gravitons from asymptotically AdS(5) space-times}",
    eprint = "hep-th/0611344",
    archivePrefix = "arXiv",
    reportNumber = "CPHT-RR019-0503",
    doi = "10.1016/j.nuclphysb.2007.02.024",
    journal = "Nucl. Phys. B",
    volume = "772",
    pages = "67--102",
    year = "2007"
}

@article{Polyakov:2018zvc,
    author = "Polyakov, Maxim V. and Schweitzer, Peter",
    title = "{Forces inside hadrons: pressure, surface tension, mechanical radius, and all that}",
    eprint = "1805.06596",
    archivePrefix = "arXiv",
    primaryClass = "hep-ph",
    doi = "10.1142/S0217751X18300259",
    journal = "Int. J. Mod. Phys. A",
    volume = "33",
    number = "26",
    pages = "1830025",
    year = "2018"
}

@article{Panteleeva:2023evj,
    author = "Panteleeva, J. Yu. and Epelbaum, E. and Gegelia, J. and Mei{\ss}ner, U. -G.",
    title = "{Electromagnetic and gravitational local spatial densities for spin-1 systems}",
    eprint = "2305.01491",
    archivePrefix = "arXiv",
    primaryClass = "hep-ph",
    doi = "10.1007/JHEP07(2023)237",
    journal = "JHEP",
    volume = "07",
    pages = "237",
    year = "2023"
}

@article{Freese:2022yur,
    author = "Freese, Adam and Cosyn, Wim",
    title = "{Spatial densities of momentum and forces in spin-one hadrons}",
    eprint = "2207.10787",
    archivePrefix = "arXiv",
    primaryClass = "hep-ph",
    reportNumber = "NT@UW-2206",
    doi = "10.1103/PhysRevD.106.114013",
    journal = "Phys. Rev. D",
    volume = "106",
    number = "11",
    pages = "114013",
    year = "2022"
}

@article{Lorce:2019sbq,
    author = "Lorc{\'e}, C{\'e}dric and Lowdon, Peter",
    title = "{Universality of the Poincar{\'e} gravitational form factor constraints}",
    eprint = "1908.02567",
    archivePrefix = "arXiv",
    primaryClass = "hep-th",
    doi = "10.1140/epjc/s10052-020-7779-z",
    journal = "Eur. Phys. J. C",
    volume = "80",
    number = "3",
    pages = "207",
    year = "2020"
}

@article{Cosyn:2019aio,
    author = "Cosyn, Wim and Cotogno, Sabrina and Freese, Adam and Lorc{\'e}, C{\'e}dric",
    title = "{The energy-momentum tensor of spin-1 hadrons: formalism}",
    eprint = "1903.00408",
    archivePrefix = "arXiv",
    primaryClass = "hep-ph",
    doi = "10.1140/epjc/s10052-019-6981-3",
    journal = "Eur. Phys. J. C",
    volume = "79",
    number = "6",
    pages = "476",
    year = "2019"
}

@article{Sun:2020wfo,
    author = "Sun, Bao-Dong and Dong, Yu-Bing",
    title = "{Gravitational form factors of $\rho$ meson with a light-cone constituent quark model}",
    eprint = "2002.02648",
    archivePrefix = "arXiv",
    primaryClass = "hep-ph",
    doi = "10.1103/PhysRevD.101.096008",
    journal = "Phys. Rev. D",
    volume = "101",
    number = "9",
    pages = "096008",
    year = "2020"
}

@article{Freese:2019bhb,
    author = {Freese, Adam and Clo{\"e}t, Ian C.},
    title = "{Gravitational form factors of light mesons}",
    eprint = "1903.09222",
    archivePrefix = "arXiv",
    primaryClass = "nucl-th",
    doi = "10.1103/PhysRevC.100.015201",
    journal = "Phys. Rev. C",
    volume = "100",
    number = "1",
    pages = "015201",
    year = "2019",
    note = "[Erratum: Phys.Rev.C 105, 059901 (2022)]"
}

@article{Allahverdiyeva:2023fhn,
    author = "Allahverdiyeva, Minaya and Mamedov, Shahin",
    title = "{Vector meson gravitational form factors and generalized parton distributions at finite temperature within the soft-wall AdS/QCD model}",
    eprint = "2302.03383",
    archivePrefix = "arXiv",
    primaryClass = "hep-ph",
    doi = "10.1140/epjc/s10052-023-11607-7",
    journal = "Eur. Phys. J. C",
    volume = "83",
    number = "5",
    pages = "447",
    year = "2023"
}

@article{Holstein:2006ud,
    author = "Holstein, Barry R.",
    title = "{Metric modifications for a massive spin 1 particle}",
    eprint = "gr-qc/0607051",
    archivePrefix = "arXiv",
    doi = "10.1103/PhysRevD.74.084030",
    journal = "Phys. Rev. D",
    volume = "74",
    pages = "084030",
    year = "2006"
}

@article{Ji:1996ek,
    author = "Ji, Xiang-Dong",
    title = "{Gauge-Invariant Decomposition of Nucleon Spin}",
    eprint = "hep-ph/9603249",
    archivePrefix = "arXiv",
    reportNumber = "MIT-CTP-2517",
    doi = "10.1103/PhysRevLett.78.610",
    journal = "Phys. Rev. Lett.",
    volume = "78",
    pages = "610--613",
    year = "1997"
}

@article{Cotogno:2019xcl,
    author = "Cotogno, Sabrina and Lorc{\'e}, C{\'e}dric and Lowdon, Peter",
    title = "{Poincar{\'e} constraints on the gravitational form factors for massive states with arbitrary spin}",
    eprint = "1905.11969",
    archivePrefix = "arXiv",
    primaryClass = "hep-th",
    doi = "10.1103/PhysRevD.100.045003",
    journal = "Phys. Rev. D",
    volume = "100",
    number = "4",
    pages = "045003",
    year = "2019"
}

@unpublished{Hippelainen:2026pion,
    author = {Hippel\"ainen, Antti and Jokela, Niko and J\"arvinen, Matti},
    title = "{Gravitational form factors of pions in V-QCD}",
    note = "in preparation",
    year = "2026, in preparation"
}

@article{DeWolfe:1999cp,
    author = "DeWolfe, O. and Freedman, D. Z. and Gubser, S. S. and Karch, A.",
    title = "{Modeling the fifth-dimension with scalars and gravity}",
    eprint = "hep-th/9909134",
    archivePrefix = "arXiv",
    reportNumber = "HUTP-99-A048, MIT-CTP-2903",
    doi = "10.1103/PhysRevD.62.046008",
    journal = "Phys. Rev. D",
    volume = "62",
    pages = "046008",
    year = "2000"
}

@article{Hoyos:2019kzt,
    author = "Hoyos, Carlos and Jokela, Niko and Logares, Daniel",
    title = "{Scattering length from holographic duality}",
    eprint = "1910.13929",
    archivePrefix = "arXiv",
    primaryClass = "hep-th",
    reportNumber = "HIP-2019-35/TH",
    doi = "10.1103/PhysRevD.101.046028",
    journal = "Phys. Rev. D",
    volume = "101",
    number = "4",
    pages = "046028",
    year = "2020"
}

@article{Hoyos:2020fjx,
    author = "Hoyos, Carlos and Jokela, Niko and Logares, Daniel",
    title = "{Scattering length in holographic confining theories}",
    eprint = "2005.06904",
    archivePrefix = "arXiv",
    primaryClass = "hep-th",
    doi = "10.1103/PhysRevD.102.086006",
    journal = "Phys. Rev. D",
    volume = "102",
    number = "8",
    pages = "086006",
    year = "2020"
}

@article{Hoyos:2022ptd,
    author = "Hoyos, Carlos and Jokela, Niko and Logares, Daniel",
    title = "{Revisiting the chiral effective action in holographic models}",
    eprint = "2203.05916",
    archivePrefix = "arXiv",
    primaryClass = "hep-th",
    reportNumber = "HIP-2022-4/TH",
    doi = "10.1103/PhysRevD.107.026017",
    journal = "Phys. Rev. D",
    volume = "107",
    number = "2",
    pages = "026017",
    year = "2023"
}

@article{Shi:2023oll,
    author = "Shi, Chao and Li, Jicheng and Yin, Pei-Lin and Jia, Wenbao",
    title = "{Unpolarized generalized parton distributions of light and heavy vector mesons}",
    eprint = "2302.02388",
    archivePrefix = "arXiv",
    primaryClass = "hep-ph",
    doi = "10.1103/PhysRevD.107.074009",
    journal = "Phys. Rev. D",
    volume = "107",
    number = "7",
    pages = "074009",
    year = "2023"
}

@article{Neubelt:2019sou,
    author = "Neubelt, Matt J. and Sampino, Andrew and Hudson, Jonathan and Tezgin, Kemal and Schweitzer, Peter",
    title = "{Energy momentum tensor and the D-term in the bag model}",
    eprint = "1911.08906",
    archivePrefix = "arXiv",
    primaryClass = "hep-ph",
    doi = "10.1103/PhysRevD.101.034013",
    journal = "Phys. Rev. D",
    volume = "101",
    number = "3",
    pages = "034013",
    year = "2020"
}

@article{Best:1997qp,
    author = "Best, C. and Gockeler, M. and Horsley, R. and Ilgenfritz, Ernst-Michael and Perlt, H. and Rakow, Paul E. L. and Schafer, A. and Schierholz, G. and Schiller, A. and Schramm, S.",
    title = "{Pion and rho structure functions from lattice QCD}",
    eprint = "hep-lat/9703014",
    archivePrefix = "arXiv",
    reportNumber = "DESY-97-041, DESY-97-41, HUP-EP-97-15, UFTP-440-1997",
    doi = "10.1103/PhysRevD.56.2743",
    journal = "Phys. Rev. D",
    volume = "56",
    pages = "2743--2754",
    year = "1997"
}

@article{Stegeman:2025tdl,
    author = "Stegeman, Roy and Zwicky, Roman",
    title = "{Gluon gravitational D-form factor: the {\ensuremath{\sigma}}-meson as a dilaton confronted with lattice data II}",
    eprint = "2512.12315",
    archivePrefix = "arXiv",
    primaryClass = "hep-ph",
    doi = "10.1007/JHEP05(2026)159",
    journal = "JHEP",
    volume = "05",
    pages = "159",
    year = "2026"
}

@article{Polyakov:2002yz,
    author = "Polyakov, M. V.",
    title = "{Generalized parton distributions and strong forces inside nucleons and nuclei}",
    eprint = "hep-ph/0210165",
    archivePrefix = "arXiv",
    reportNumber = "RUB-TP2-14-02",
    doi = "10.1016/S0370-2693(03)00036-4",
    journal = "Phys. Lett. B",
    volume = "555",
    pages = "57--62",
    year = "2003"
}

\end{document}